\documentclass[sigplan,screen,nonacm]{acmart}

\usepackage[page]{appendix}
\usepackage{algorithm2e} %
\usepackage{arydshln}
\usepackage{balance}
\usepackage{booktabs}
\usepackage{xspace} %
\usepackage[compatibility=false]{caption}
\usepackage{comment}
\usepackage{enumitem}
\makeatletter
\def\mdseries@tt{m}
\@namedef{ver@lineno.sty}{9999/12/31}
\@namedef{opt@lineno.sty}{}
\makeatother
\usepackage[newfloat,frozencache=true]{minted}
\usepackage{multirow}
\usepackage{pifont}
\newcommand{\cmark}{\ding{51}}

\newcommand{\xmark}{\ding{55}}
\usepackage{subcaption}

\hypersetup{
colorlinks = true,
linkcolor = black,
citecolor = black,
}
\newcommand{\ie}{i.e.,~}
\newcommand{\eg}{e.g.,~}

\newcommand{\Cite}[1]{~\cite{#1}}

\newcommand{\fakeitem}{\vspace*{2pt} \noindent $\bullet$ \hskip .01in}

\newcommand{\PROPNAME}{CALIFORMS}
\newcommand{\Propname}{Califorms}
\newcommand{\propname}{califorms}
\newcommand{\prop}{caliform}

\newcommand{\proped}{califormed}
\newcommand{\proping}{califorming}

\newcommand{\bytename}{security byte}
\newcommand{\Instname}{CFORM}
\newcommand{\INSTNAME}{\texttt{CFORM}}
\newcommand{\cstruct}{struct}
\newcommand{\cclass}{class}
\newcommand{\verbose}[1]{\texttt{#1}}

\setlist[itemize]
{leftmargin=8pt,
 itemindent=0pt,
 topsep=2pt,
 partopsep=2pt,
 parsep=2pt,
 itemsep=2pt,
 listparindent=\parindent{}
}

\begin{document}

\title{Practical Byte-Granular Memory Blacklisting using \Propname{}}

\author{Hiroshi Sasaki}
\affiliation{%
  \institution{Columbia University}}
\email{sasaki@cs.columbia.edu}

\author{Miguel A. Arroyo}
\affiliation{%
  \institution{Columbia University}}
\email{miguel@cs.columbia.edu}

\author{M. Tarek Ibn Ziad}
\affiliation{%
  \institution{Columbia University}}
\email{mtarek@cs.columbia.edu}

\author{Koustubha Bhat$^\dagger{}$}
\affiliation{%
  \institution{Vrije Universiteit Amsterdam}}
\email{k.bhat@vu.nl}\thanks{${}^\dagger{}$Part of this work was carried out while the author was a visiting student at Columbia University.}

\author{Kanad Sinha}
\affiliation{%
  \institution{Columbia University}}
\email{kanad@cs.columbia.edu}

\author{Simha Sethumadhavan}
\affiliation{%
  \institution{Columbia University}}
\email{simha@cs.columbia.edu}

\renewcommand{\shortauthors}{H. Sasaki, M. Arroyo, M. Tarek Ibn Ziad, K. Bhat, K. Sinha, and S. Sethumadhavan}

\begin{abstract}

Recent rapid strides in memory safety tools and hardware have improved software quality and security.
While coarse-grained memory safety has improved, achieving memory safety at the
granularity of individual objects remains a challenge due to high performance
overheads which can be between $\sim$1.7x$-$2.2x.
In this paper, we present a novel idea called \emph{\Propname}{}, and associated
program observations, to obtain a low overhead security solution for practical,
byte-granular memory safety.

The idea we build on is called memory blacklisting, which prohibits
a program from accessing certain memory regions based on program
semantics.  State of the art hardware-supported memory blacklisting
while much faster than software blacklisting creates memory fragmentation
(of the order of few bytes) for each use of the blacklisted location.
In this paper, we observe that metadata used for blacklisting
can be stored in dead spaces in a program's data memory and that
this metadata can be integrated into microarchitecture by changing
the cache line format.  Using these observations, \Propname{} based
system proposed in this paper reduces the performance overheads of
memory safety to $\sim$1.02x$-$1.16x while providing byte-granular
protection and maintaining very low hardware overheads.

The low overhead offered by \Propname{} enables always on, memory safety for
small and large objects alike, and the fundamental idea of storing metadata in
empty spaces, and microarchitecture can be used for other security and
performance applications.

\end{abstract}

\date{}
\settopmatter{printfolios=true}
\maketitle
\thispagestyle{firstpage}

\section{Introduction}\label{sec:introduction}

With recent interest in microarchitecture side channels, it is important not to lose sight of more traditional software security threats.
Security is a full-system property where both software and hardware have to be secure for a system to be secure.
Historically, program memory safety violations have provided a significant opportunity for exploitation: for instance, a recent report from Microsoft revealed that the root cause of more than half of all exploits were software memory safety violations\Cite{Blackhat:2016wi}.
In response to the severity of this threat, improvements in software checking tools, such as AddressSanitizer\Cite{Serebryany:2012wl}, and advances in the form of commercial hardware support for memory safety such as Oracle's ADI\Cite{ADI:2015ha} and Intel's MPX\Cite{Oleksenko:2018kz} have enabled programmers to detect and fix memory safety violations before deploying software.

Current software and hardware-supported solutions excel at providing coarse-grained memory safety, \ie{}detecting memory access beyond arrays and \verbose{malloc}'d regions (struct and class instances).
However, they are not suitable for fine-grained memory safety (\ie{}detecting overflows within objects, such as fields within a struct, or members within a class) due to the high performance overheads and/or need for making intrusive changes to the source code\Cite{Song:2018tm}.
For instance, a recent work that aims to provide intra-object overflow protection functionality incurs a 2.2x performance overhead\Cite{Duck:2018jy}.
These overheads are problematic because they not only reduce the number of
pre-deployment tests that can be performed, but also impede post-deployment continuous monitoring, which researchers have pointed out is necessary for detecting benign and malicious memory safety violations\Cite{Serebryany:2018wg}.
Thus, a low overhead memory safety solution that can enable continuous
monitoring and provide complete program safety has been elusive.

The source of overheads stem from how current designs store and use metadata necessary for enforcing memory safety.
In Intel MPX\Cite{Oleksenko:2018kz}, Hardbound\Cite{Devietti:2008kw}, CHERI\Cite{Woodruff:2014tn,Woodruff:2019jk}, and PUMP\Cite{Dhawan:2015kv}, the metadata is stored for each pointer, and each data or code memory access through a pointer performs checks using the metadata.
Since C/C++ memory accesses tend to be highly pointer based, the performance and energy overheads of accessing metadata can be significant in such systems.
Furthermore, the management of metadata especially if it is stored in a disjoint manner from the pointer can also create significant engineering complexity in terms of performance and usability.
This was evidenced by the fact that compilers like LLVM and GCC dropped support for Intel MPX in their mainline after an initial push to integrate into the toolchain\Cite{Oleksenko:2018kz}.

Our approach for reducing overheads is two-fold.  First, instead of
checking access bounds for each pointer access, we blacklist all memory
locations that should never be accessed. In theory, this is a strictly
weaker form of security than whitelisting but we argue that in practice,
blacklisting can be more practical because of its ease of deployment and
low overheads. Informally, deployments apply whitelisting techniques
partially to reduce overheads and be backward compatible which reduces
their security, while blacklisting techniques can be applied more
broadly due to their low overheads. Additionally, blacklisting
techniques complement defenses in existing systems better since they do
not require intrusive changes.

Our second optimization is the novel metadata storage scheme.
We observe that by using dead memory spaces in the program, we can store
metadata needed for memory safety for free for nearly half of the program objects.
These dead spaces occur because of language alignment requirements and are inserted by the compiler.
When we cannot find a naturally occurring dead space, we manually insert a dead space.
The overhead due to this dead space is smaller than traditional methods for
storing metadata because of how we represent the metadata: our metadata is smaller (one byte) as opposed to multiple bytes with traditional whitelisting or blacklisting memory safety techniques.

A natural question is how the dead (more commonly referred to as \textit{padding}) bytes can be distinguished from normal bytes in memory.
A straightforward scheme results in one bit of additional storage per byte to identify if a byte is a dead byte; this scheme results in a space overhead of 12.5\%.
We reduce this overhead to one bit per 64B cache line (0.2\% overhead) without any loss of precision by only reformatting how data is stored in cache lines.
Our technique, \emph{\Propname{}}, uses one bit of additional storage to identify if the cache line associated with the memory contains any dead bytes.
For \emph{\proped{}} cache lines, \ie{}lines which contain dead bytes, the actual data is stored following the ``header'', which indicates the location of dead bytes, as shown in \autoref{fig:califorms-overview}.

With this support, it is easy to describe how a \Propname{} based system for memory safety works.
The dead bytes, either naturally harvested or manually inserted, are used to indicate memory regions that should never be accessed by a program (\ie{}blacklisting).
If an attacker accesses these regions, we detect this rogue access without any additional metadata accesses as our metadata resides inline.

\setcounter{figure}{0} %
\begin{figure}[!t] %
  \centering
  \includegraphics[width=0.84\linewidth]{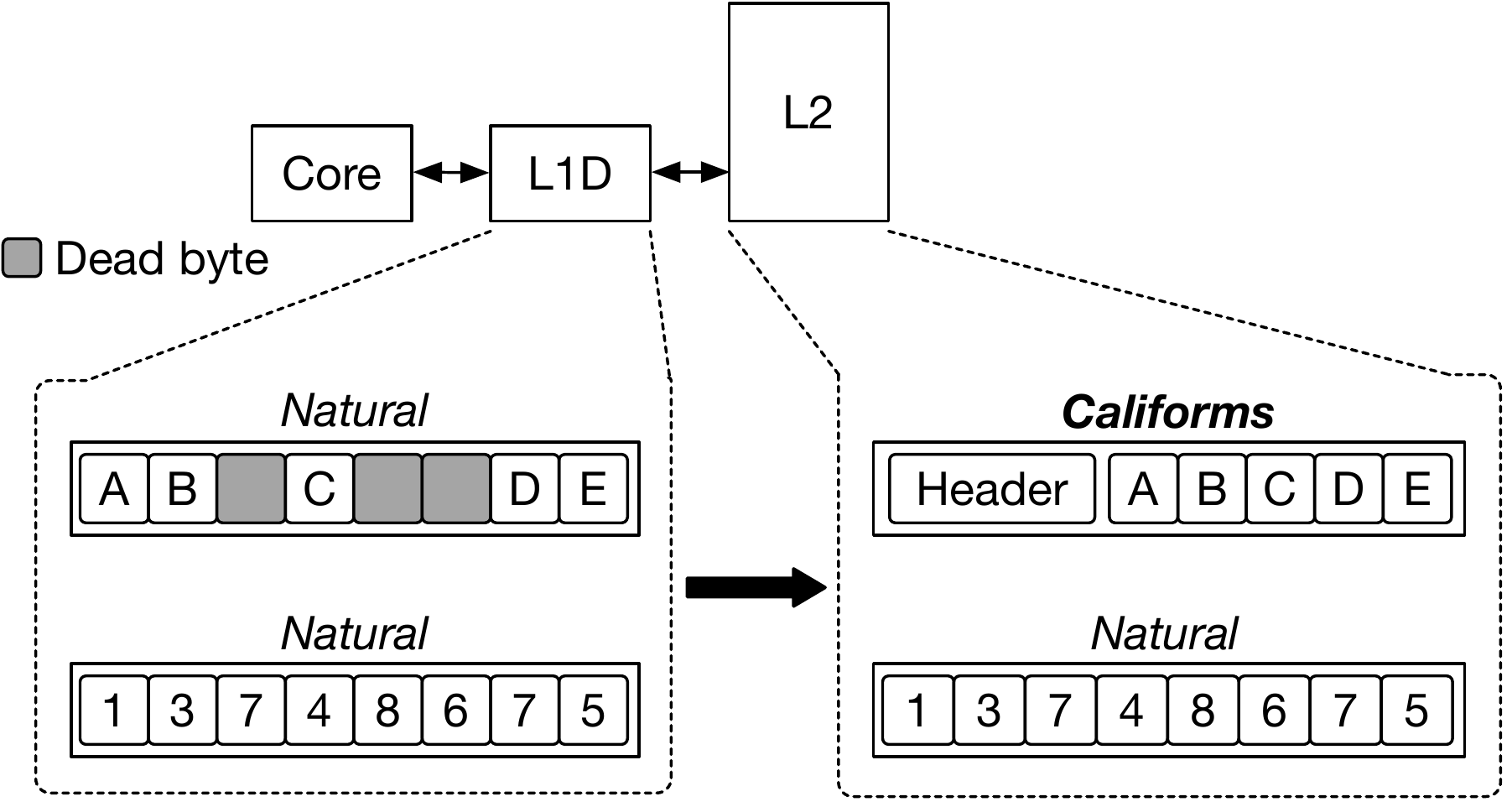}
  \caption{\Propname{} offers memory safety by detecting accesses to dead bytes in memory. Dead bytes are not stored beyond the L1 data cache and identified using a special header in the L2 cache (and beyond) resulting in very low overhead. The conversion between these formats happens when lines are filled or spilled between the L1 and L2 caches. The absence of dead bytes results in the cache lines stored in the same natural format across memory system.}\label{fig:califorms-overview}
\end{figure} %

Our experimental results on the SPEC CPU2006 benchmark suite indicate that the overheads of \Propname{} are quite low: software overheads range from 2 to 14\% slowdown (or alternatively, 1.02x to 1.16x performance overhead) depending on the amount and location of padding bytes used.
This provides the functionality for the user/customer to tune the security according to their performance requirements.
Hardware induced overheads are also negligible, on average less than 1\%.
All of the software transformations are performed using the LLVM compiler framework using a front-end source-to-source transformation.
These overheads are substantially lower compared to the state-of-the-art software or hardware supported schemes (viz., 2.2x performance and 1.1x memory overheads for EffectiveSan\Cite{Duck:2018jy}, and 1.7x performance and 2.1x memory overheads for Intel MPX\Cite{Oleksenko:2018kz}).

\section{Motivation}\label{sec:motivation}

\setcounter{figure}{1} %
\setcounter{subfigure}{0}
\begin{listing*}[!t] %
  \begin{minipage}[t]{0.15\textwidth}
    \centering
    \begin{minted}
    [
    framesep=2mm,
    baselinestretch=1.0,
    fontsize=\scriptsize,
    style=bw,
    stripnl=false,
    ]
    {cpp}
struct A {
  char c;
  int i;
  char buf[64];
  void (*fp)();
  double d;
}








    \end{minted}
    \captionof{subfigure}{Original.}\label{lst:policy-original}
  \end{minipage}
  \begin{minipage}[t]{0.25\textwidth}
    \centering
    \begin{minted}
    [
    framesep=2mm,
    baselinestretch=1.0,
    fontsize=\scriptsize,
    style=bw,
    stripnl=false,
    highlightlines={5},
    highlightcolor=lightgray,
    ]
    {cpp}
struct A_opportunistic {
  char c;
  /* compiler inserts padding
   * bytes for alignment */
  char padding_bytes[3];
  int i;
  char buf[64];
  void (*fp)();
  double d;
}





    \end{minted}
    \captionof{subfigure}{Opportunistic.}\label{lst:policy-opportunistic}
  \end{minipage}
  \hspace{1mm}
  \begin{minipage}[t]{0.28\textwidth}
    \centering
    \begin{minted}
    [
    framesep=2mm,
    baselinestretch=1.0,
    fontsize=\scriptsize,
    style=bw,
    stripnl=false,
    highlightlines={4,6,8,10,12,14},
    highlightcolor=lightgray,
    ]
    {cpp}
struct A_full {
  /* we protect every field with
   * random security bytes */
  char security_bytes[2];
  char c;
  char security_bytes[1];
  int i;
  char security_bytes[3];
  char buf[64];
  char security_bytes[2];
  void (*fp)();
  char security_bytes[1];
  double d;
  char security_bytes[2];
}
    \end{minted}
    \captionof{subfigure}{Full.}\label{lst:policy-full}
  \end{minipage}
  \hspace{1mm}
  \begin{minipage}[t]{0.28\textwidth}
    \centering
    \begin{minted}
    [
    framesep=2mm,
    baselinestretch=1.0,
    fontsize=\scriptsize,
    style=bw,
    stripnl=false,
    highlightlines={7,9,11},
    highlightcolor=lightgray,
    ]
    {cpp}
struct A_intelligent {
  char c;
  int i;
  /* we protect boundaries
   * of arrays and pointers with
   * random security bytes */
  char security_bytes[3];
  char buf[64];
  char security_bytes[2];
  void (*fp)();
  char security_bytes[3];
  double d;
}


    \end{minted}
    \captionof{subfigure}{Intelligent.}\label{lst:policy-intelligent}
  \end{minipage}
  \captionof{listing}{Example of three \bytename{}s harvesting strategies: (b) \textit{opportunistic} uses the existing padding bytes as \bytename{}s, (c) \textit{full} protect every field within the struct with \bytename{}s, and (d) \textit{intelligent} surrounds arrays and pointers with \bytename{}s.}\label{lst:policies}
\end{listing*} %

One of the key ways in which we mitigate the overheads for fine-grained memory safety is by opportunistically harvesting padding bytes in programs to store metadata.
So how often do these occur in programs?
Before we answer that question let us concretely understand padding bytes with an example.
Consider the \texttt{struct~A} defined in \autoref{lst:policy-original}.
Let us say the compiler inserts a three-byte padding in between \texttt{char~c} and \texttt{int~i} as in \autoref{lst:policy-opportunistic} because of the C language requirement that integers should be padded to their natural size (which we assume to be four bytes here).
These types of paddings are not limited to C/C++ but also many other languages and their runtime implementations.
To obtain a quantitative estimate on the amount of paddings, we developed a compiler pass to statically collect the padding size information.
\autoref{fig:struct-density} presents the histogram of struct densities for SPEC CPU2006 C and C++ benchmarks and the V8 JavaScript engine.
Struct density is defined as the sum of the size of each field divided by the total size of the \cstruct{} including the padding bytes (\ie{}the smaller or sparse the struct density the more padding bytes the struct has).
The results reveal that $45.7\%$ and $41.0\%$ of structs within SPEC and V8, respectively, have at least one byte of padding.
This is encouraging since even without introducing additional padding bytes (no memory overhead), we can offer protection for certain compound data types restricting the remaining attack surface.

\setcounter{figure}{2} %
\begin{figure}[!t] %
  \centering
  \subcaptionbox{SPEC CPU2006 C and C++ benchmarks.\label{fig:struct-density-speccpu2006}}{\includegraphics[width=0.48\columnwidth]{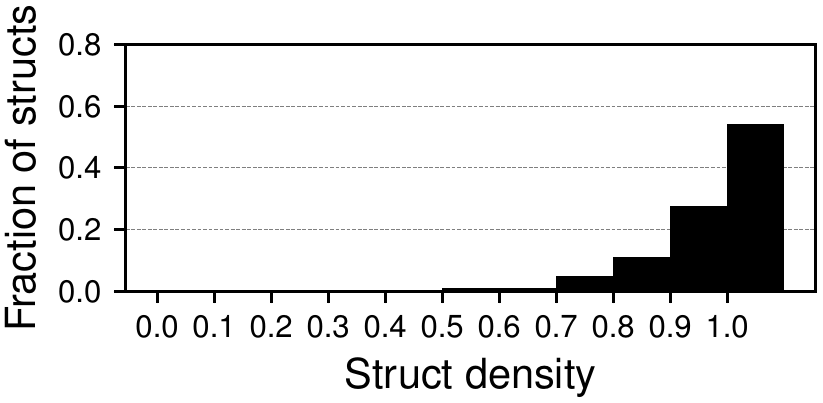}}
  \subcaptionbox{V8 JavaScript engine.\label{fig:struct-density-v8}}{\includegraphics[width=0.48\columnwidth]{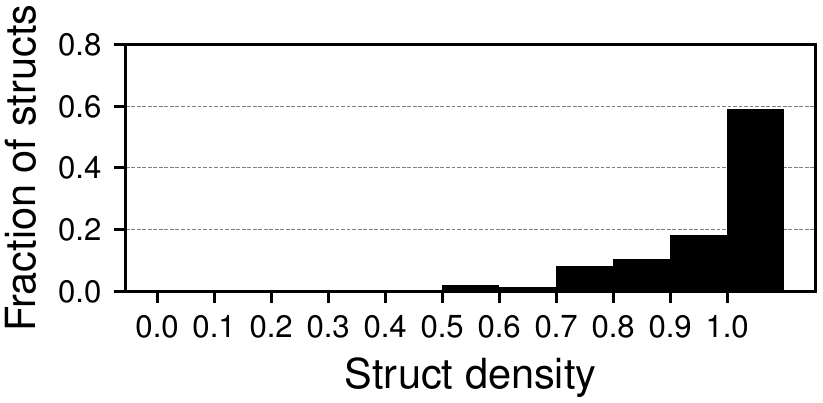}}
  \caption{Struct density histogram of SPEC CPU2006 benchmarks and the V8 JavaScript engine. More than 40\% of the structs have at least one padding byte.}\label{fig:struct-density}
\end{figure} %

\begin{figure}[!t] %
  \centering
  \includegraphics[width=0.60\linewidth]{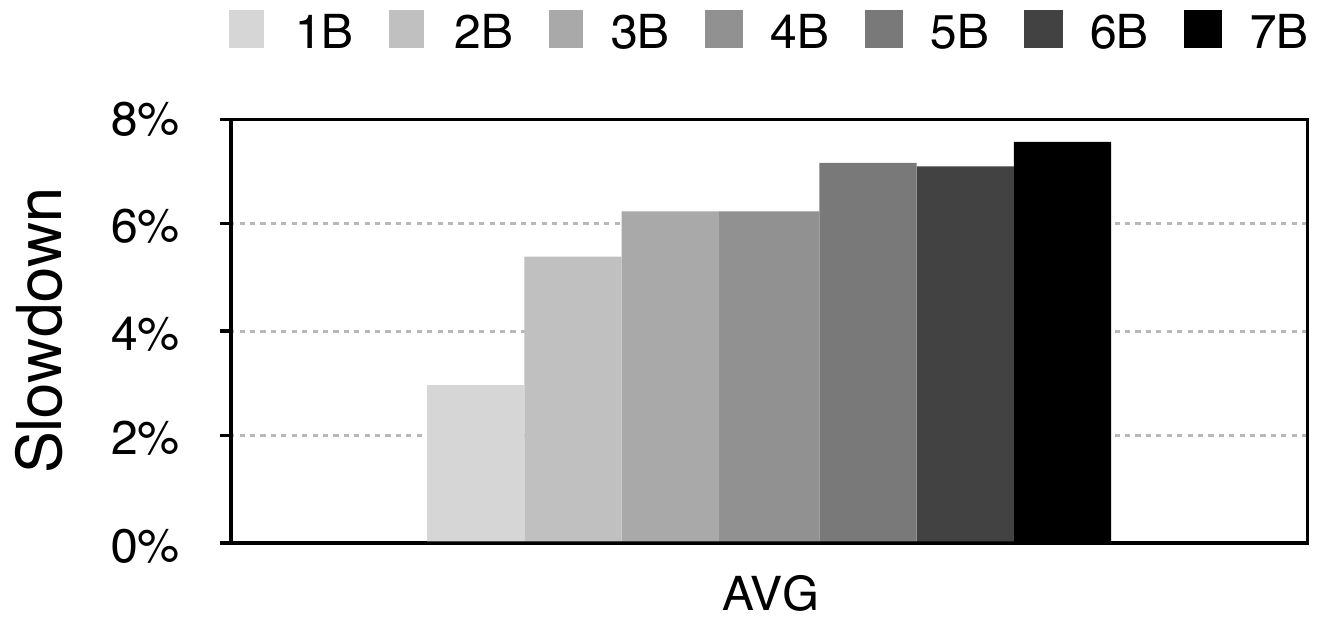}
  \caption{Average performance overhead with additional paddings (one byte to seven bytes) inserted for every field within structs (and classes) of SPEC CPU2006 C and C++ benchmarks.}\label{fig:avg-perf-fixed-full-speccpu2006}
\end{figure} %

Naturally, one might inquire about the safety for the rest of the program.
To offer protection for all defined compound data types (called the full strategy), we can insert random sized padding bytes, also referred to as \bytename{}s, between every field of a \cstruct{} or member of a \cclass{} as in \autoref{lst:policy-full}.
Random sized \bytename{}s are chosen to provide a probabilistic defense as fixed sized \bytename{}s can be jumped over by an attacker once s/he identifies the actual size (and the exact memory layout).
Additionally, by carefully choosing the minimum and maximum sizes for insertion, we can keep the average \bytename{} size small (such as two or three bytes).
Intuitively, the higher the unpredictability (or randomness) there is within the memory layout, the higher the security level we can offer.

While the full strategy provides the widest coverage, not all of the \bytename{}s provide the same security utility.
For example, basic data types such as \texttt{char} and \texttt{int} cannot be easily overflowed past their bounds.
The idea behind the intelligent insertion strategy is to prioritize insertion of \bytename{}s into security-critical locations as presented in \autoref{lst:policy-intelligent}.
We choose data types which are most prone to abuse by an attacker via overflow type accesses: (1) arrays and (2) data and function pointers.
In the example in \autoref{lst:policy-intelligent}, the array \texttt{buf[64]} and the function pointer \texttt{fp} are protected with random sized \bytename{}s.
While it is possible to utilize padding bytes present between other data types without incurring memory overheads, doing so would come at an additional performance overhead.

In comparison to opportunistic harvesting, the other more secure strategies (\eg{}full strategy) come at an additional performance overhead.
We analyze the performance trend in order to decide how many \bytename{}s can be reasonably inserted.
For this purpose we developed an LLVM pass which pads every field of a \cstruct{} with fixed size paddings.
We measure the performance of SPEC CPU2006 benchmarks by varying the padding size from one byte to seven bytes.
The detailed evaluation environment and methodology is described later in \autoref{sec:evaluation}.

\autoref{fig:avg-perf-fixed-full-speccpu2006} demonstrates the average slowdown when inserting additional bytes for harvesting.
As expected, we can see the performance overheads increase as we increase the padding size, mainly due to ineffective cache usage.
On average the slowdown is $3.0\%$ for one byte and $7.6\%$ for seven bytes of padding.
The figure presents the ideal (lower bound) performance overhead when fully inserting \bytename{}s into compound data types; the hardware and software modifications we introduce add additional overheads on top of these numbers.
We strive to provide a mechanism that allows the user to tune the security level at the cost of performance and thus explore several \bytename{} insertion strategies to reduce the performance overhead in the paper.

\section{Full System Overview}\label{sec:overview}

The \Propname{} framework consists of multiple components we discuss in the following sections:

\fakeitem\textbf{Architecture Support.} An ISA extension of a machine instruction called \INSTNAME{} that performs \proping{} (\ie{}(un)setting \bytename{}s) of cache lines, and a privileged \Propname{} exception which is raised upon misuse of \bytename{}s (\autoref{sec:architecture}). %

\fakeitem\textbf{Microarchitecture Design.} New cache line formats that enable low cost access to the metadata --- we propose different \Propname{} for L1 cache vs. L2 cache and beyond (\autoref{sec:microarchitecture}).

\fakeitem\textbf{Software Design.} Compiler, memory allocator and operating system extensions which insert the \bytename{}s at compile time and manages the \bytename{}s via the \INSTNAME{} instruction at runtime (\autoref{sec:software}).

At compile time each compound data type, a \cstruct{} or a \cclass{}, is examined and \bytename{}s are added according to a user defined insertion policy viz.\ opportunistic, full or intelligent, by a source-to-source translation pass.
When we run the binary with \bytename{}s, when compound data type instances are created in the heap dynamically, we use a new version of \verbose{malloc} that issues \INSTNAME{} instructions to set the \bytename{}s after the space is allocated.
When the \INSTNAME{} instruction is executed, the cache line format is transformed at the L1 cache controller (assuming a cache miss) and is inserted into the L1 data cache.
Upon an L1 eviction, the L1 cache controller re-\propname{} the cache line to meet the \Propname{} of the L2 cache.

While we add additional metadata storage to the caches, we refrain from doing so for main memory and persistent storage to keep the changes local within the CPU core.
When a \proped{} cache line is evicted from the last-level cache to main memory,
we keep the cache line \proped{} and store the additional one metadata bit into
spare ECC bits similar to Oracle's ADI\Cite{ADI:2015ha}.\footnote{ADI stores four bits of metadata per cache line for allocation granularity enforcement while \Propname{} stores one bit for sub-allocation granularity enforcement.}
When a page is swapped out from main memory, the page fault handler stores the metadata for all the cache lines within the page into a reserved address space managed by the operating system; the metadata is reclaimed upon swap in.
Therefore, our design keeps the cache line format \proped{} throughout the memory hierarchy.
A \proped{} cache line is un-\proped{} only when the corresponding bytes cross the boundary where the \proped{} data cannot be understood by the other end, such as writing to I/O (\eg{}pipe, filesystem or network socket).
Finally, when an object is freed, the freed bytes are \proped{} and zeroed for offering temporal safety.

At runtime, when a rogue load or store accesses a \proped{} byte the hardware returns a privileged, precise security exception to the next privilege level which can take any appropriate action including terminating the program.

\section{Architecture Support}\label{sec:architecture}

\subsection{\Instname{} Instruction}

\begin{table}[t] %
  \makeatletter
  \newcommand*{\textoverline}[1]{$\overline{\hbox{#1}}\m@th$}
  \makeatother
  \centering
  \caption{K-map for the \INSTNAME{} instruction. X represents ``Don't Care''.}\label{tab:cform}
  \scalebox{0.80}{
  \begingroup
  \begin{tabular}{lcccc}
                                                                             & \multicolumn{1}{l}{}  & \multicolumn{3}{c}{R2, R3} \\[0.3em]
                                                                             &                       & X, \textoverline{Allow}          & \textoverline{Set}, Allow        & Set, Allow \\[0.1em] \cmidrule{3-5}
    \multicolumn{1}{l}{\multirow{2}{*}{\rotatebox[origin=c]{90}{Initial}}}   & Regular Byte          & \multicolumn{1}{c}{Regular Byte} & \multicolumn{1}{c}{Exception} &  \multicolumn{1}{c}{Security Byte} \\ \cmidrule{3-5}
    \multicolumn{1}{l}{}                                                     & Security Byte         & \multicolumn{1}{c}{Security Byte}& Regular Byte & \multicolumn{1}{c}{Exception} \\ \cmidrule{3-5}
  \end{tabular}
  \endgroup
  }
  \vspace{1em}
\end{table} %

The format of the instruction is ``\INSTNAME{} \texttt{R1}, \texttt{R2}, \texttt{R3}''.
The value in register \texttt{R1} points to the starting (cache aligned) address in the virtual address space, denoting the start of the 64B chunk which fits in a single 64B cache line.
\autoref{tab:cform} represents a K-map for the \INSTNAME{} instruction.
The value in register \texttt{R2} indicates the attributes of said region represented in a bit vector format ($1$ to set and $0$ to unset the \bytename{}).
The value in register \texttt{R3} is a mask to the corresponding 64B region, where $1$ allows and $0$ disallows changing the state of the corresponding byte.
The mask is used to perform partial updates of metadata within a cache line.
We throw a privileged \Propname{} exception when the \INSTNAME{} instruction tries to set a \bytename{} to an existing \bytename{} location, and unset a \bytename{} from a normal byte.

The \INSTNAME{} instruction is treated similar to a store instruction in the processor pipeline, where it first fetches the corresponding cache line into the L1 data cache upon an L1 miss (assuming a write allocate cache policy).
Next, it manipulates the bits in the metadata storage to appropriately set or unset the \bytename{}s.\footnote{We also investigate the possibility of using a variant of \INSTNAME{} instruction which does not store the modified cache line into the L1 data cache, just like the non-temporal (or streaming) load/store instructions (\eg{}\texttt{MOVNTI}, \texttt{MOVNTQ}, etc) in \autoref{subsec:allocation_and_deallocation}.}

\subsection{Privileged Exceptions}

When the hardware detects an access violation, it throws a privileged exception once the instruction becomes non-speculative.
There are some library functions which violate the aforementioned operations \bytename{}s such as \mbox{\texttt{memcpy}} so we need a way to suppress the exceptions.
In order to whitelist such functions, we manipulate the exception mask registers and let the exception handler decide whether to suppress the exception or not.
Although privileged exception handling is more expensive than handling user-level exceptions (because it requires a context switch to the kernel), we stick with the former to limit the attack surface. We rely on the fact that the exception itself is a rare event and would have negligible effect on performance.

\section{Microarchitecture Design}\label{sec:microarchitecture}

The microarchitectural support for our technique aims to keep the common case fast: L1 cache uses the straightforward scheme of having one bit of additional storage per byte.
All \proped{} cache lines are transformed to the straightforward scheme at the L1 data cache controller so that typical loads and stores which hit in the L1 cache do not have to perform address calculations to figure out the location of original data (which is required for \Propname{} of L2 cache and beyond).
This design decision guarantees that for the common case the latencies will not be affected due to security functionality.
Beyond the L1, the data is stored in the optimized \proped{} format, \ie{}one bit of additional storage for the entire cache line.
The transformation happens when the data is filled in or spilled from the L1 data cache (between the L1 and L2), and adds minimal latency to the L1 miss latency.
For main memory, we store the additional bit per cache line size in the DRAM ECC spare bits, thus completely removing any cycle time impact on DRAM access or modifications to the DIMM architecture.

\subsection{L1 Cache: Bit Vector Approach}\label{subsec:califorms_bitvector}

\begin{figure}[!t] %
  \centering
  \includegraphics[width=0.38\textwidth]{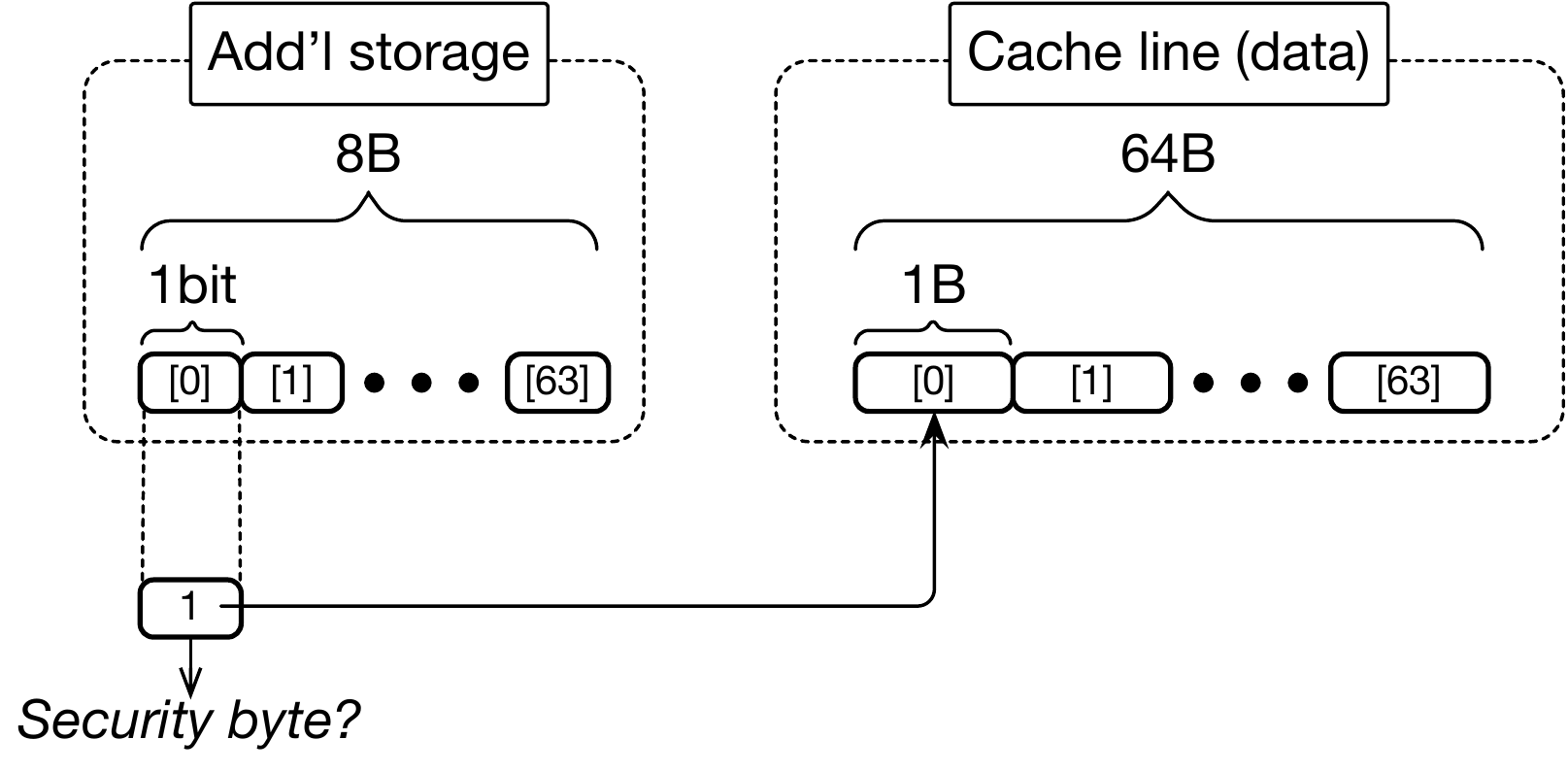}
  \caption{\Propname{}-bitvector: L1 \Propname{} implementation using a bit vector that indicates whether each byte is a \bytename{}. HW overhead of 8B per 64B cache line.}\label{fig:naive-bitvector-8byte}
\end{figure} %

To satisfy the L1 design goal we consider a naive (but low latency) approach which uses a bit vector to identify which bytes are \bytename{}s in a cache line.
Each bit of the bit vector corresponds to each byte of the cache line and represent its state (normal byte or \bytename{}).
\autoref{fig:naive-bitvector-8byte} presents a schematic view of this implementation \emph{\propname{}-bitvector}.
The bit vector requires a 64-bit (8B) bit vector per 64B cache line which adds
12.5\% storage overhead for just the L1-D caches (comparable to ECC overhead for
reliability).

\begin{figure}[!t] %
  \centering
  \includegraphics[width=0.48\textwidth]{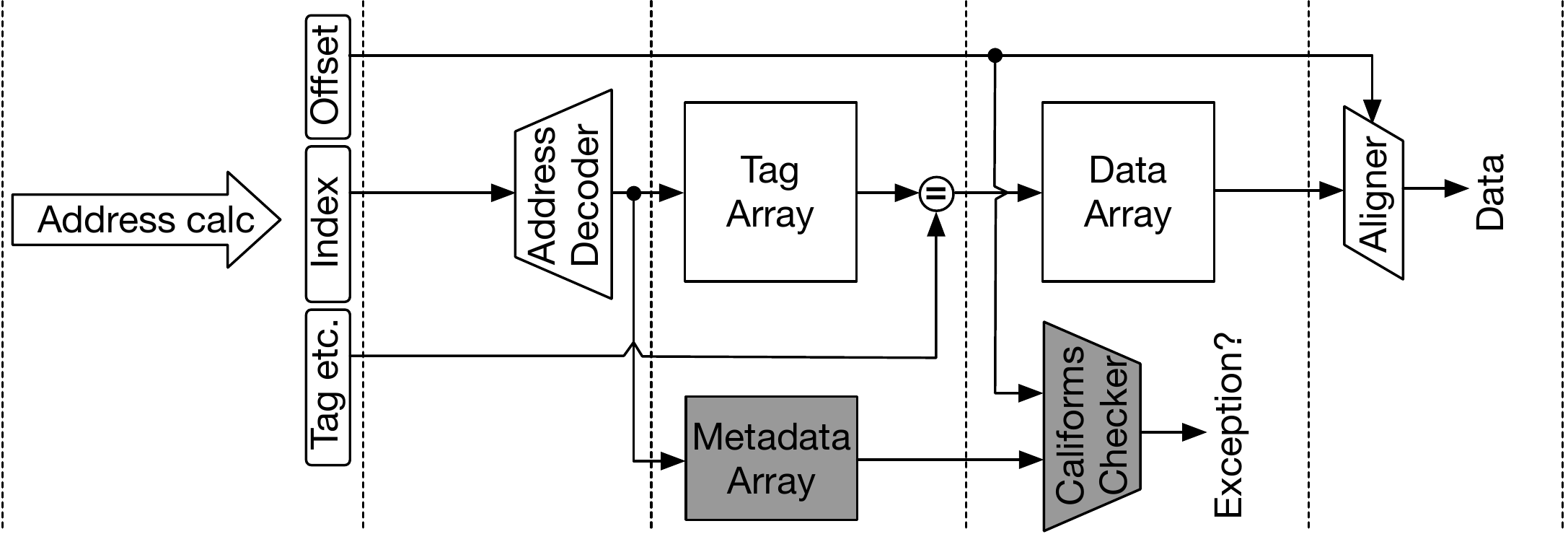}
  \caption{Pipeline diagram for the L1 cache hit operation. The shaded components correspond to \Propname{}.}\label{fig:pipeline}
\end{figure} %

\autoref{fig:pipeline} shows the L1 data cache hit path modifications for \Propname{}.
If a load accesses a \proped{} byte (which is determined by reading the bit vector) an exception is recorded to be processed when the load is ready to be committed.
Meanwhile, the load returns a pre-determined value for the security byte (in our design the value 0 which is the value that the memory region is initialized to upon deallocation).
The reason to return the pre-determined value is to avoid a speculative side channel attack to identify security byte locations and is discussed in greater detail in \autoref{sec:discussion}.
On store accesses to \proped{} bytes we report an exception before the store commits.

\subsection{L2 Cache and Beyond: Sentinel Approach}

\begin{figure}[!t] %
  \centering
  \includegraphics[width=0.44\textwidth]{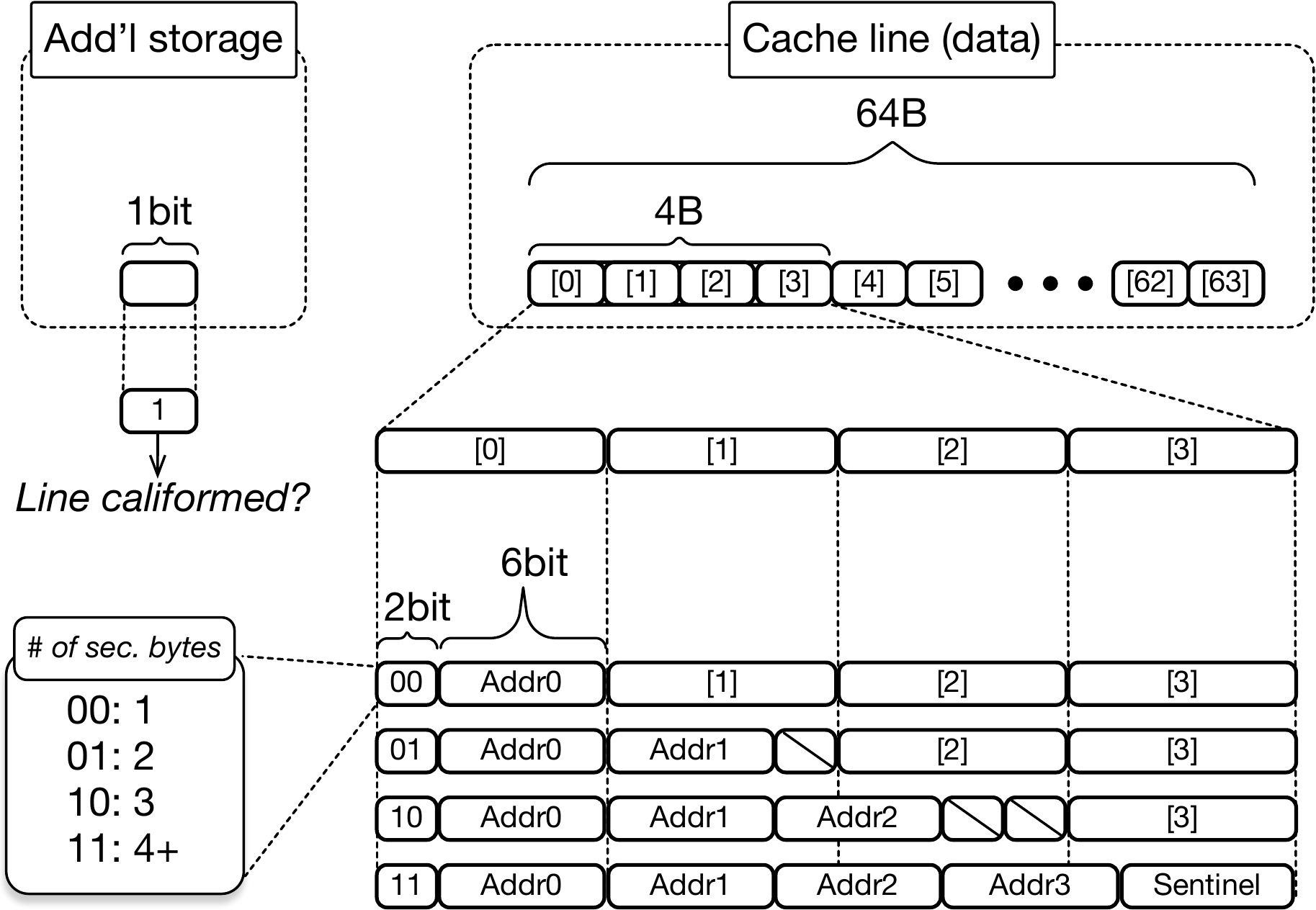}
  \caption{\Propname{}-sentinel that stores a bit vector in \bytename{} locations. HW overhead of 1-bit per 64B cache line.}\label{fig:califorms_sentinel_1bit}
\end{figure} %

For L2 and beyond, we take a different approach that allows us to recognize whether each byte is a \bytename{} with fewer bits, as using the L1 metadata format throughout the system will increase the cache area overhead by 12.5\%, which may not be acceptable.
\autoref{fig:califorms_sentinel_1bit} illustrates our proposed \emph{\propname{}-sentinel}, which has a 1-bit or 0.2\% metadata overhead per 64B cache line.

The key insight that enables these savings is the following observation: the number of addressable bytes in a cache line is less than what can be represented by a single byte (we only need six bits).
For example, let us assume that there is (at least) one security byte in a 64B cache line.
Considering a byte granular protection there are at most 63 unique values (bytes) that non-security bytes can have.
Therefore, we are guaranteed to find a six bit pattern which is not present in any of the normal bytes' least (or most) significant six bits.
We use this pattern as a sentinel value to represent the security bytes within the cache line.

If we store the six bit sentinel value as additional metadata, the overhead will be seven bits (six bits plus one bit to specify if the cache line is \proped{}) per cache line.
Instead, we propose a new cache line format which stores the sentinel value within a \bytename{} to reduce the metadata overhead down to one bit per cache line.
The idea is to use four different formats depending on the number of \bytename{}s in the cache line, as we explain below.

\Propname{}-sentinel stores the metadata into the first four bytes (at most) of the 64B cache line.
Two bits of the 0th byte is used to specify the number of \bytename{}s within the cache line: \texttt{00}, \texttt{01}, \texttt{10} and \texttt{11} represent one, two, three, and four or more \bytename{}s, respectively.
If there is only one \bytename{} in the cache line, we use the remaining six bits of the 0th byte to specify the location of the \bytename{}
(and the original value of the 0th byte is stored in the \bytename{}).
Similarly when there is two or three \bytename{}s in the cache line, we use the bits of the 1st and 2nd bytes to locate them.
The key observation is that, we gain two bits per \bytename{} since we only need six bits to specify a location in the cache line.
Therefore when we have four \bytename{}s, we can locate four addresses and have six bits remaining in the first four bytes.
This remaining six bits can be used to store a sentinel value, which allows us to have any number of additional \bytename{}s.

\begin{figure}[!t] %
    \centering
    \begin{minted}
    [
    framesep=2mm,
    baselinestretch=1.0,
    fontsize=\scriptsize,
    style=bw,
    stripnl=false,
    highlightlines=,
    highlightcolor=lightgray,
    escapeinside=||,
    ]
    {text}
 1: Read the |\Propname{}| metadata for the evicted line and OR them
 2: |\bf{if}| result is 0 |\bf{then}|
 3:    Evict the line as is and set |\Propname{}| bit to 0
 4: |\bf{else}|
 5:    Set |\Propname{}| bit to 1
 6:    Perform following operations on the cache line:
 7:       Scan least 6-bit of every byte to determine sentinel
 8:       Get locations of 1st 4 |\bytename{}s|
 9:       Store data of 1st 4 bytes in locations obtained in 8:
10:       Fill the 1st 4 bytes based on |\autoref{fig:califorms_sentinel_1bit}|
11:       Use the sentinel to mark the remaining |\bytename{}s|
12: |\bf{end}|
    \end{minted}
    \captionof{algocf}{\Propname{} conversion from the L1 cache (\propname{}-bitvector) to L2 cache (\propname{}-sentinel).}\label{alg:bitvector-to-sentinel}
\end{figure} %

\begin{figure}[!t] %
    \centering
    \begin{minted}
    [
    framesep=2mm,
    baselinestretch=1.0,
    fontsize=\scriptsize,
    style=bw,
    stripnl=false,
    highlightlines=,
    highlightcolor=lightgray,
    escapeinside=||,
    ]
    {text}
 1: Read the |\Propname{}| bit for the inserted line
 2: |\bf{if}| result is 0 |\bf{then}|
 3:    Set the |\Propname{}| metadata bit vector to [0]
 4: |\bf{else}|
 5:    Perform following operations on the cache line:
 6:       Check the least significant 2-bit of byte 0
 7:       Set the metadata of byte[Addr[0-3]] to 1 based on 6:
 8:       Set the metadata of byte[Addr[byte == sentinel]] to 1
 9:       Set the data of byte[0-3] to byte[Addr[0-3]]
10:       Set the new locations of byte[Addr[0-3]] to zero
11: |\bf{end}|
    \end{minted}
    \captionof{algocf}{\Propname{} conversion from the L2 cache (\propname{}-sentinel) to L1 cache (\propname{}-bitvector).}\label{alg:sentinel-to-bitvector}
\end{figure} %

Although the sentinel value depends on the actual values within the 64B cache line, it works naturally with a write-allocate L1 cache, which is the most commonly used cache allocation policy in modern microprocessors.
The cache line format can be converted upon L1 cache eviction and insertion (\propname{}-bitvector to/from \propname{}-sentinel), and the sentinel value only needs to be found upon L1 cache eviction.
Also, it is important to note that \propname{}-sentinel supports critical-word first delivery since the \bytename{} locations can be quickly retrieved by scanning only the first 4B of the first 16B flit.
Algorithms~\ref{alg:bitvector-to-sentinel} and~\ref{alg:sentinel-to-bitvector} describe the high-level process used for converting from L1 to L2 \Propname{} and vice versa.

{
\renewcommand{\figureautorefname}{Algorithm}
\begin{figure*}[!t]
  \centering
  \includegraphics[width=1.00\linewidth]{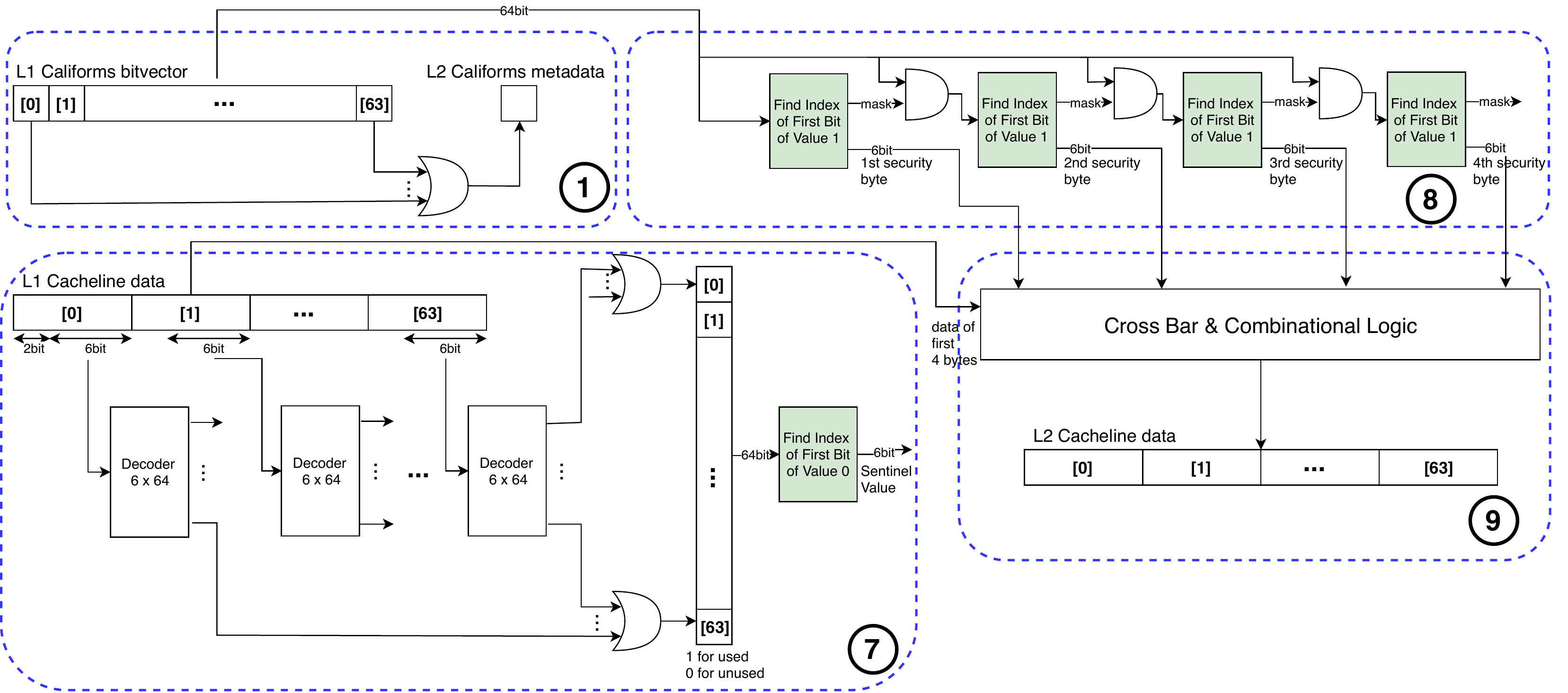}
  \caption{Logic diagram for \Propname{} conversion from the L1 cache (\propname{}-bitvector) to L2 cache (\propname{}-sentinel). The green Find-index blocks are constructed using 64 shift blocks followed by a single comparator. The circled numbers refer to the corresponding steps in \autoref{alg:bitvector-to-sentinel}.}\label{fig:bitvector-to-sentinel}
\end{figure*}
}

\autoref{fig:bitvector-to-sentinel} shows the logic diagram for the spill module.
{
\renewcommand{\figureautorefname}{Algorithm}
The circled numbers refer to the corresponding steps in \autoref{alg:bitvector-to-sentinel}.
}
In the top-left corner, the \Propname{} metadata for the evicted line is ORed to construct the L2 cache (\propname{}-sentinel) metadata bit.
The bottom-right square details the process of determining sentinel.
We scan least 6-bit of every byte, decode them, and OR the output to construct a used-values vector.
The used-values vector is then processed by a Find-index block to get the sentinel (line~7).
The Find-index block takes a 64-bit input vector and searches for the index of the first zero.
It is constructed using 64 shift blocks followed by a single comparator.

The top-right corner of \autoref{fig:bitvector-to-sentinel} shows the logic for getting the locations of the first four \bytename{}s (line~8).
It consists of four successive combinational Find-index blocks (each detecting one \bytename{}) in our evaluated design.
This logic can be easily pipelined into four stages, if needed, to completely hide the latency of the spill process in the pipeline.
Finally, we store the data of the first four bytes in locations obtained from the Find-index blocks and fill the same four bytes based on \autoref{fig:califorms_sentinel_1bit}.

{
\renewcommand{\figureautorefname}{Algorithm}
\begin{figure}[!t]
  \centering
  \includegraphics[width=0.45\textwidth]{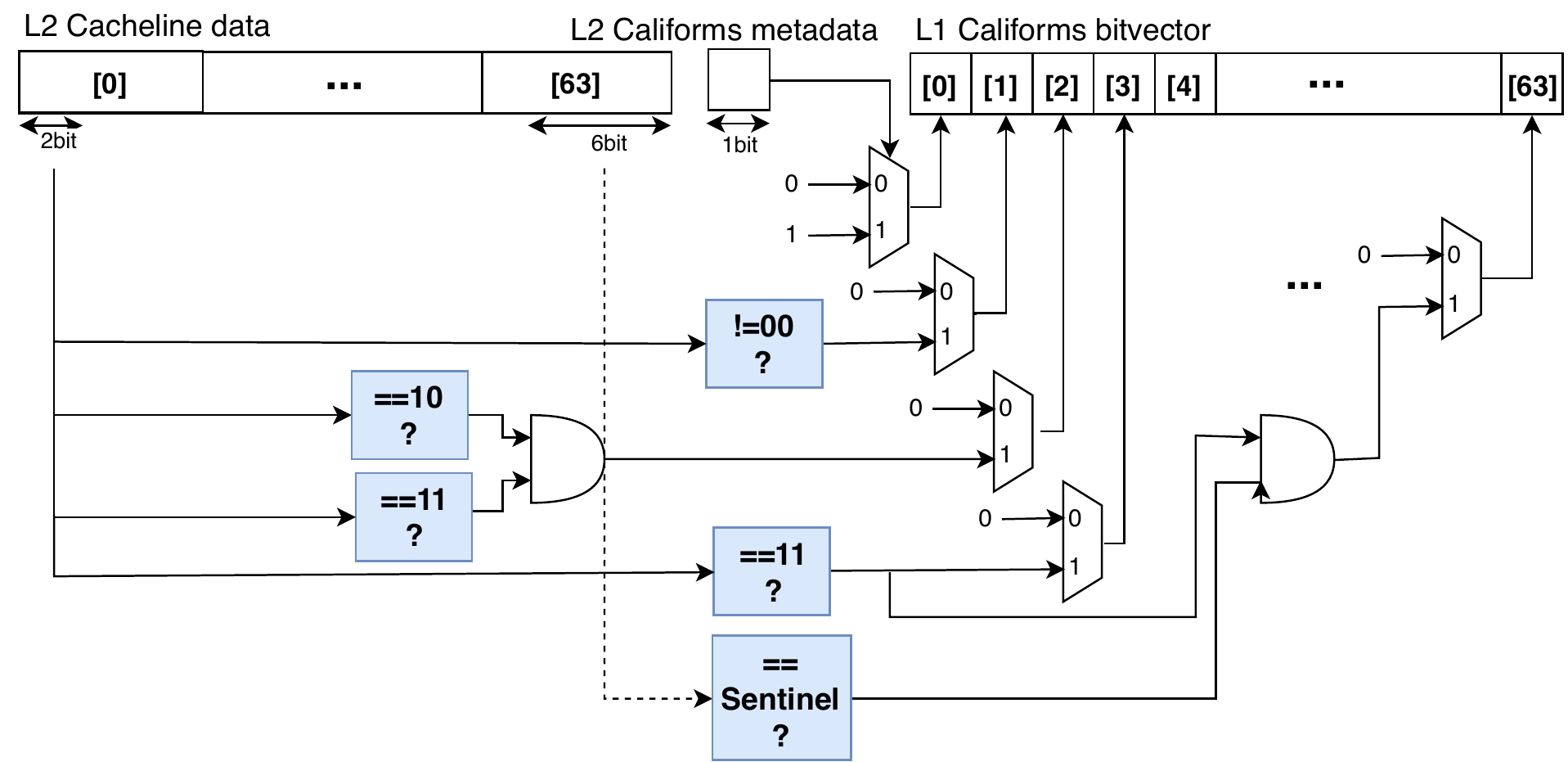}
  \caption{Logic diagram for \Propname{} conversion from the L2 cache (\propname{}-sentinel) to L1 cache (\propname{}-bitvector), as described in \autoref{alg:sentinel-to-bitvector}. The blue (==) blocks are constructed using logic comparators.}\label{fig:sentinel-to-bitvector}
\end{figure}
}

\autoref{fig:sentinel-to-bitvector} shows the logic diagram for the fill module,
{as summarized in
\renewcommand{\figureautorefname}{Algorithm}
\autoref{alg:sentinel-to-bitvector}.
}
The blue (==) blocks are constructed using logic comparators.
The \Propname{} bit of the L2 inserted line is used to control the value of the L1 cache (\propname{}-bitvector) metadata.
The first two bits of the L2 inserted line are used as inputs for the comparators to decide on the metadata bits of the first four bytes as specified in \autoref{fig:califorms_sentinel_1bit}.
Only if those two bits are \texttt{11}, the sentinel value is read from the fourth byte and fed, with the least 6-bits of each byte, to 60 comparators simultaneously to set the rest of the L1 metadata bits.
Such parallelization reduces the latency impact of the fill process.

\subsection{Load/Store Queue Modifications}

Since the \INSTNAME{} instruction updates the architecture state (writes values), it is functionally a store instruction and handled as such in the pipeline.
However, there is a key difference: unlike a store instruction, the \INSTNAME{} instruction should not forward the value to a younger load instruction whose address matches within the load/store queue~(LSQ) but instead return the value zero.
This functionality is required to provide tamper-resistance against side-channel attacks.
Additionally, upon an address match, both load and store instructions subsequent to an in flight \INSTNAME{} instruction are marked for \Propname{} exception (which is thrown when the instruction is committed).

In order to detect an address match in the LSQ with a \INSTNAME{} instruction, first a cache line address should be matched with all the younger instructions.
Subsequently upon a match, the value stored in the LSQ for the \INSTNAME{} instruction, which contains the mask value indicating to-be-\proped{} bytes, is used to confirm the final match.
To facilitate a match with a \INSTNAME{} instruction, each LSQ entry should be associated with a bit to indicate whether the entry contains a \INSTNAME{} instruction.
Detecting a complete match may take multiple cycles, however, a legitimate load/store instruction should never be forwarded a value from a \INSTNAME{} instruction, and thus the store-to-load forwarding from a \INSTNAME{} instruction is not on the critical path of the program (\ie{}its latency should not affect the performance, and we do not evaluate its effect in our evaluation).
Alternately, if LSQ modifications are to be avoided, the \INSTNAME{} instructions can be surrounded by memory serializing instructions (\ie{}ensure that \INSTNAME{} instructions are the only in flight memory instructions).

\section{Software Design}\label{sec:software}

We describe compiler support, the memory allocator changes and the operating
system changes to support \Propname{} in the following.

\subsection{Dynamic Memory Management}\label{subsec:allocation_and_deallocation}

We can consider two approaches to applying \bytename{}s: (1) \emph{Dirty-before-use}.
Unallocated memory has no \bytename{}s.
We set \bytename{}s upon allocation and unset them upon deallocation; or (2) \emph{Clean-before-use}.
Unallocated memory remains filled with \bytename{}s all the time.
We clear the \bytename{}s (in legitimate data locations) upon allocation and set them upon deallocation.
Ensuring temporal memory safety in the heap remains a non-trivial problem\Cite{Blackhat:2016wi}.
We therefore choose to follow a \emph{clean-before-use} approach in the heap, so that deallocated memory regions remain protected by \proped{} \bytename{}s\footnote{It is natural to use the non-temporal \INSTNAME{} instruction when deallocating a memory region; deallocated region is not meant to be used by the program and thus polluting the L1 data cache with those memory is harmful and should be avoided. Not evaluated in this paper is the use of non-temporal instructions which should provide better performance.}.
Additionally, in order to provide temporal memory safety, we do not reallocate recently freed regions until the heap is sufficiently consumed (quarantining).
Compared to the heap, the security benefits are limited for the stack since temporal attacks on the stack (\eg{}use-after-return attacks) are much rarer.
Hence, we apply the \emph{dirty-before-use} scheme on the stack.

\subsection{Compiler}\label{subsec:object_type_layouts}

Our compiler-based instrumentation infers where to place \bytename{}s within target objects, based on their type layout information.
The compiler pass supports three insertion policies: the first \emph{opportunistic} policy supports \bytename{}s insertion into existing padding bytes within the objects, and the other two support modifying object layouts to introduce randomly sized \bytename{} spans that follow the \emph{full} or \emph{intelligent} strategies described in \autoref{sec:motivation}.
The first policy aims at retaining interoperability with external code modules (\eg{}shared libraries) by avoiding type layout modification.
Where this is not a concern, the latter two policies help offer stronger security coverage --- exhibiting a tradeoff between security and performance.

\noindent

\subsection{Operating System Support}\label{subsec:softwaresupport}

We need the following support in the operating system:

\fakeitem\textbf{Privileged Exceptions.}
As the \Propname{} exception is privileged, the operating system needs to properly handle it as with other privileged exceptions (\eg{}page faults).
We also assume the faulting address is passed in an existing register so that it can be used for reporting/investigation purposes.
Additionally, for the sake of usability and backwards compatibility, we have to accommodate copying operations similar in nature to \texttt{memcpy}.
For example, a simple \cstruct{} to \cstruct{} assignment could trigger this behavior, thus leading to a potential breakdown of \proped{} software.
Hence, in order to maintain usability, we allow whitelisting functionality to suppress the exceptions.
This is done by issuing a privileged store instruction to modify the value of exception mask registers before entering and after exiting the according piece of code.
We discuss the implications of this design choice in \autoref{sec:discussion}.

\fakeitem\textbf{Page Swaps.}
As we have discussed in \autoref{sec:overview}, data with \bytename{}s is stored in main memory in a \proped{} format.
When a page with \proped{} data is swapped out from main memory, the page fault handler needs to store the metadata for the entire page into a reserved address space managed by the operating system; the metadata is reclaimed upon swap in.
The kernel has enough address space in practice (kernel's virtual address space is 128TB for a 64-bit Linux with 48-bit virtual address space) to store the metadata for all the processes on the system since the size of the metadata for a 4KB page consumes only 8B.

\section{Security Discussion}\label{sec:discussion}

\subsection{Threat Model}\label{subsec:threatmodel}

For the security evaluation of this work, we assume a threat model comparable to that used in contemporary related works.
We assume the victim program to have one or more vulnerabilities that an attacker can exploit to gain arbitrary read and write capabilities in the memory.
Furthermore, we assume that the adversary has access to the source code of the program, therefore s/he is able to glean all source-level information and/or deterministic compilation results from it (\eg{}find code gadgets within the program and determine non-\proped{} layouts of data structures).
However, s/he does not have access to the host binary (\eg{}server-side applications).
Finally, we assume that all hardware is trusted: it does not contain and/or is not subject to bugs arising from exploits such as physical or glitching attacks.
Due to its recent rise in relevance however, we maintain side channel attacks in our design of \Propname{} within the purview of our threats.
Specifically, we accommodate attack vectors seeking to leak the location and value of \bytename{}s.

\subsection{Hardware Attacks and Mitigations}

\fakeitem\textbf{Metadata Tampering Attacks.} A key feature of \Propname{} as a metadata-based safety mechanism is the absence of programmer visible metadata in the general case (apart from a metadata bit in the page information maintained by higher privilege software).
Beyond the implications for its storage overhead, this also means that our technique is immune to attacks that explicitly aim to leak or tamper the metadata to bypass the respective defense.
This, in turn, implies a smaller attack surface so far as software maintenance of metadata is concerned.

\fakeitem\textbf{Bit-granularity Attacks.} \Propname{}'s capability of fine-grained memory protection is the key enabler for intra-object overflow detection.
However, our byte granular mechanism is not enough for protecting bit-fields without turning them into \verbose{char} bytes functionally.
This should not be a major detraction since \bytename{}s can still be added around composites of bit-fields.

\fakeitem\textbf{Heterogeneous Architectural Attacks.} \Propname{}' hardware modifications affect the memory hierarchy.
Hence, its protection is lost whenever one of its layers is bypassed (\eg{}heterogeneous architectures or DMA is used).
Mitigating this requires that these mechanisms always respect the \bytename{} semantics by propagating them along the respective memory structures and detecting accesses to them.
If the algorithm used for \proping{} is used by accelerators then attacks through heterogeneous components can also be averted.

\fakeitem\textbf{Side-Channel Attacks.} Our design takes multiple steps to be resilient to side channel attacks.
Firstly, we purposefully avoid timing variances introduced due to our hardware modifications in order to avoid timing based side channel attacks.
Additionally, to avoid speculative execution side channels ala Spectre\Cite{Kocher:2018wl}, our design returns zero on a load to a \bytename{}, thus preventing speculative disclosure of metadata.
We augment this further by requiring that deallocated objects (heap or stack) be zeroed out in software\Cite{Milburn:2017fn}.
This is to avoid the following attack scenario: consider a case if the attacker somehow knows that the padding locations should contain a non-zero value (for instance, because s/he knows the object allocated at the same location prior to the current object had non-zero values).
However, while speculatively disclosing memory contents of the object, s/he discovers that the padding location contains a zero instead.
As such, s/he can infer that the padding there contains a \bytename{}.
If deallocations were accompanied with zeroing, however, this assumption does not hold.

\subsection{Software Attacks and Mitigations}

\fakeitem\textbf{Coverage-Based Attacks.} For \proping{} the padding bytes (in an object), we need to know the precise type information of the allocated object.
This is not always possible in C-style programs where \texttt{void*} allocations may be used.
In these cases, the compiler may not be able to infer the correct type, in which case intra-object support may be skipped for such allocations.
Similarly, our metadata insertion policies (viz.,~intelligent and full) require changes to the type layouts.
This means that interactions with external modules that have not been compiled with \Propname{} support may need (de)serialization to remain compatible. %
For an attacker, such points in execution may appear lucrative because of inserted \bytename{}s getting stripped away in those short periods.
We note however that the opportunistic policy can still remain in place to offer some protection.

On the other hand, for those interactions that remain oblivious to type layout modifications (\eg{}passing a pointer to an object that shall remain opaque within the external module), our hardware-based implicit checks have the benefit of persistent tampering protection, even across binary module boundaries.

\fakeitem\textbf{Whitelisting Attacks.} Our concession of allowing whitelisting of certain functions was necessary to make \Propname{} more usable in common environments without requiring significant source modifications.
However, this also creates a vulnerability window wherein an adversary can piggy back on these functions in the source to bypass our protection.
To confine this vector, we keep the number of whitelisted functions as minimal as possible.

\fakeitem\textbf{Derandomization Attacks.} Since \Propname{} can be bypassed if an attacker can guess a \bytename{}s location, it is crucial that it be placed unpredictably. For the attacker to carry out a guessing attack, s/he first needs to obtain the virtual memory address of the object they want to corrupt, and then overwrite a certain number of bytes within that object.
To know the address of the object of interest, s/he typically has to scan the process' memory: the probability of scanning without touching any of the security bytes is $(1-P/N){}^O$ where $O$ is number of allocated objects, $N$ is the size of each object, and $P$ is number of security bytes within that object.
With 10\% padding ($P/N = 0.1$), when $O$ reaches $250$, the attack success goes to $10^{-20}$.
If the attacker can somehow reduce $O$ to $1$, which represents the ideal case for the attacker, the probability of guessing the element of interest is $1/{7^n}$ (since we insert 1--7 wide \bytename{}s), compounding as the number of paddings to be guessed ($=n$) increases.

The randomness is, however, introduced statically akin to \texttt{randstruct} plugin introduced in recent Linux kernels which randomizes structure layout of those which are specified (it does not offer detection of rogue accesses unlike \Propname{} do)\Cite{LKML:randstruct,LWN:randstruct}.
The static nature of the technique may make it prone to brute force attacks like BROP\Cite{Bittau:2014jv} which repeatedly crashes the program until the correct configuration is guessed.
This could be prevented by having multiple versions of the same binary with different padding sizes or simply by better logging, when possible.
Another mitigating factor is that BROP attacks require specific type of program semantics, namely, automatic restart-after-crash with the same memory layout.
Applications with these semantics can be modified to spawn with a different padding layout in our case and yet satisfy application level requirements.

\section{Performance Evaluation}\label{sec:evaluation}

\subsection{Hardware Overheads}\label{subsec:evaluation_hardware}

\begin{table*}[t!] %
  \caption{Area, delay and power overheads of \Propname{} (GE represents gate equivalent). L1 \Propname{} (\propname{}-bitvector) adds negligible delay and power overheads to the L1 cache access.\label{tab:vlsi}}{%
  \centering
\resizebox{0.99\textwidth}{!}{
    \begin{tabular}{c  c c c  c c c  c c c  c c c}
        \toprule
  Design & \multicolumn{3}{c}{Main synthesis results} & \multicolumn{3}{c}{L1 overheads} & \multicolumn{3}{c}{Fill overheads}  & \multicolumn{3}{c}{Spill overheads} \\
    Name  & Area  (GE) & Delay  ($ns$)  & Power  ($m W$)    & Area  (\%)  & Delay(\%) & Power  (\%)   & Area  (GE) & Delay  ($ns$)  & Power  ($m W$) & Area  (GE) & Delay  ($ns$)  & Power  ($m W$) \\ %
    \midrule
Baseline & 347,329.19 & 1.62 & 15.84 & --- & --- & --- & --- & --- & --- & --- & --- & --- \\
\midrule
L1 \Propname{} & 412,263.87 & 1.65 & 16.17 & 18.69 & 1.85 & 2.12 & 8,957.16 & 1.43 & 0.18 & 34,561.80 & 5.50 & 0.52  \\
 \bottomrule
\end{tabular}}
}
\end{table*} %

\fakeitem\textbf{Cache Access Latency Impact of \Propname{}.}
\Propname{} adds additional state and operations to the L1 data cache and the interface between the L1 and L2 caches.
The goal of this section is to evaluate the access latency impact of the additional state and operations described in \autoref{sec:microarchitecture}.
Qualitatively, the metadata area overhead of L1 \Propname{} is 12.5\%, and the access latency should not be impacted as the metadata lookup can happen in parallel with the L1 tag access;
the L1 to/from L2 \propname{} conversion should also be simple enough so that its latency can be completely hidden.
However, the metadata area overhead can increase the L1 tag access latency and the conversions might add little latency.
Without loss of generality, we measure the access latency impact of adding \propname{}-bitvector on a 32KB direct mapped L1 cache in the context of a typical energy optimized tag, data, formatting L1 pipeline with multicycle fill/spill handling.
For the implementation we use the 65nm TSMC core library, and generate the SRAM arrays with the ARM Artisan memory compiler.
\autoref{tab:vlsi} summarizes the results for the L1 \Propname{} (\propname{}-bitvector).

As expected, the overheads associated with the \propname{}-bitvector are minor in terms of delay (1.85\%) and power consumption (2.12\%).
We found the SRAM area to be the dominant component in the total cache area (around $98\%$) where the overhead was 18.69\%, higher than 12.5\%.
The results of fill/spill modules are reported separately in the right hand side of \autoref{tab:vlsi}.

The latency impact of the fill operation is within the access period of the L1 design.
Thus, the \proping{} operation can be folded completely within the pipeline stages that are responsible for bringing cache lines from L2 to L1.

The timing delay of the (less performance sensitive) spill operation is larger than that of the fill operation (5.5$ns$ vs. 1.4$ns$) as we use pure combinational logic to construct the \propname{}-sentinel format in one cycle, as shown in Figure~\ref{fig:bitvector-to-sentinel}.
{
\renewcommand{\figureautorefname}{Algorithm}
This cycle period can be reduced by dividing the operations of \autoref{alg:bitvector-to-sentinel} (lines~7 to 11) into two or more pipeline stages.
}
For instance, getting the locations of the first four \bytename{}s (line~8) consists of four successive combinational blocks (each detecting one \bytename{}) in our evaluated design.
This logic can be easily pipelined into four stages.
Therefore we believe that the latency of both the fill and spill operations can be minimal (or completely hidden) in the pipeline.

\fakeitem\textbf{Performance with Additional Cache Access Latency.}
Our results from the VLSI implementation imply that there will be no additional L2/L3 latency imposed by implementing \Propname{}.
However, this might not be the case depending on several implementation details (\eg{}target clock frequency) so we pessimistically assume that the L2/L3 access latency incurs additional one cycle latency overhead.
In order to evaluate the performance of the additional latency posed by \Propname{}, we perform detailed microarchitectural simulations.

We use ZSim\Cite{Sanchez:2013fp} as the processor simulator and use PinPoints\Cite{Patil:2004jr} with Intel Pin\Cite{Luk:2005hw}, to select representative simulation regions of the SPEC CPU2006 benchmarks with \texttt{ref} inputs compiled with Clang version 6.0.0 with ``\texttt{-O3 -fno-strict-aliasing}'' flags.
We do not warmup the simulator upon executing each SimPoint region but instead use a relatively large interval length of 500M instructions to avoid any warmup issues.
We set MaxK used in SimPoint region selection to 30.\footnote{For some benchmark-input pairs we have seen discrepancies
in the number of instructions measured by PinPoints vs. ZSim and thus the appropriate SimPoint regions might not be simulated. Those inputs are: \texttt{foreman\_ref\_encoder\_main} for \texttt{h264ref}
and \texttt{pds-50} for \texttt{soplex}. Also, due to time constraints, we could not complete executing SimPoint for \texttt{h264ref} with \texttt{sss\_encoder\_main} input and excluded it from the evaluation.}

\begin{table}[t] %
  \centering
  \caption{Hardware configuration of the simulated system.}\label{tab:hw_parameters}
  \scalebox{0.80}{\begingroup
  \begin{tabular}{r l}
  \toprule
  Core                & x86-64 Intel Westmere-like OoO core at 2.27GHz \\ %
  \midrule
  L1 inst.\ cache     & 32KB, 4-way, 3-cycle latency                   \\
  \midrule
  L1 data cache       & 32KB, 8-way, 4-cycle latency                   \\
  \midrule
  L2 cache            & 256KB, 8-way, 7-cycle latency                  \\
  \midrule
  L3 cache            & 2MB, 16-way, 27-cycle latency                  \\
  \midrule
  DRAM                & 8GB, DDR3-1333                                 \\ %
  \bottomrule
  \end{tabular}
  \endgroup
  }
\end{table} %

\begin{figure}[!t] %
  \centering
  \includegraphics[width=0.99\linewidth]{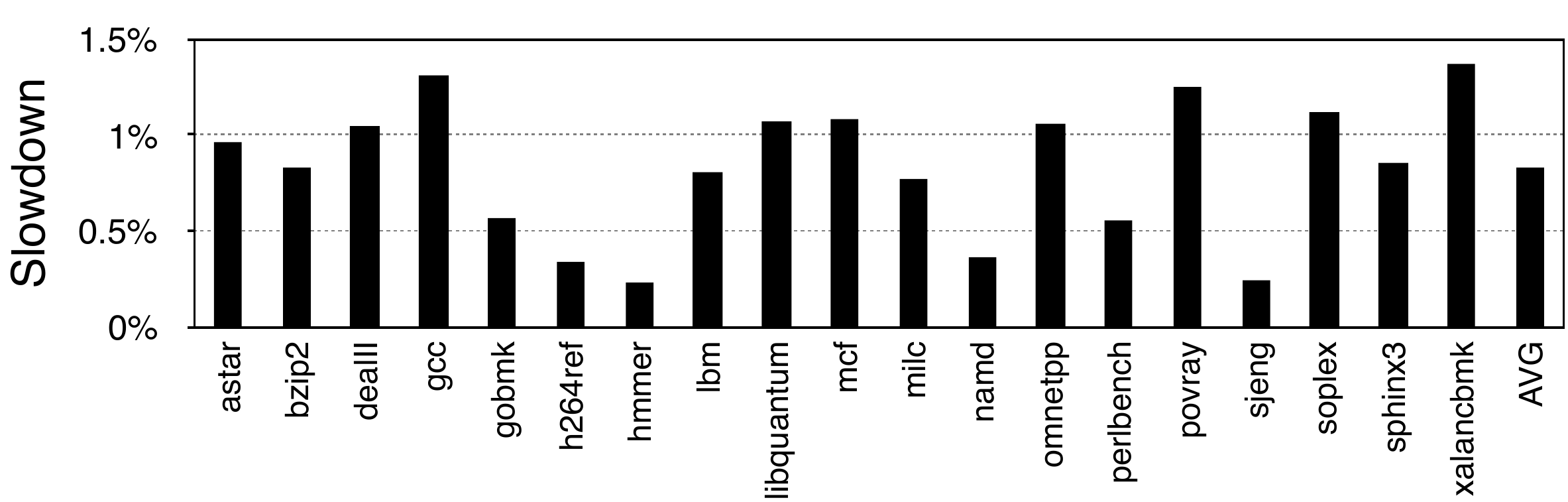}
  \caption{Slowdown with additional one-cycle access latency for both L2 and L3 caches.}\label{fig:perf-addl-latency-speccpu2006}
\end{figure} %

\autoref{tab:hw_parameters} shows the parameters of the processor, an Intel Westmere-like out-of-order core which has been validated against a real system whose performance and microarchitectural events to be commonly within 10\%\Cite{Sanchez:2013fp}.
We evaluate the performance when both L2 and L3 caches incur additional latency of one cycle.

As shown in \autoref{fig:perf-addl-latency-speccpu2006} slowdowns range from 0.24\% (\texttt{hmmer}) to 1.37\% (\texttt{xalancbmk}).
The average performance slowdown is 0.83\% which is negligible and is well in the range of error when executed on real systems.

\subsection{Software Performance Overheads}\label{subsec:evaluation_software}

\begin{figure*}[!t] %
  \centering
  \includegraphics[width=1.00\linewidth]{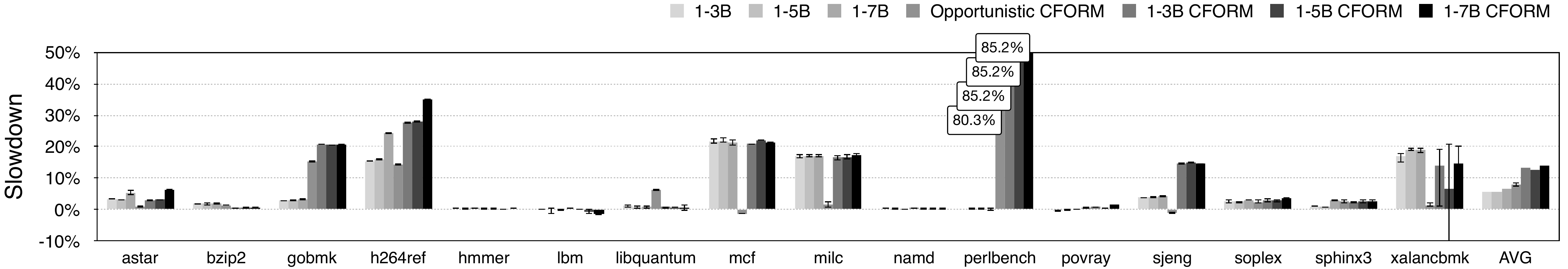}
  \caption{Slowdown of the opportunistic policy, and full insertion policy with random sized \bytename{}s (with and without \INSTNAME{} instructions). The average slowdowns of opportunistic and full insertion policies are 6.2\% and 14.2\%, respectively.}\label{fig:perf-random-full-speccpu2006}
\end{figure*} %

\begin{figure*}[!t] %
  \centering
  \includegraphics[width=1.00\linewidth]{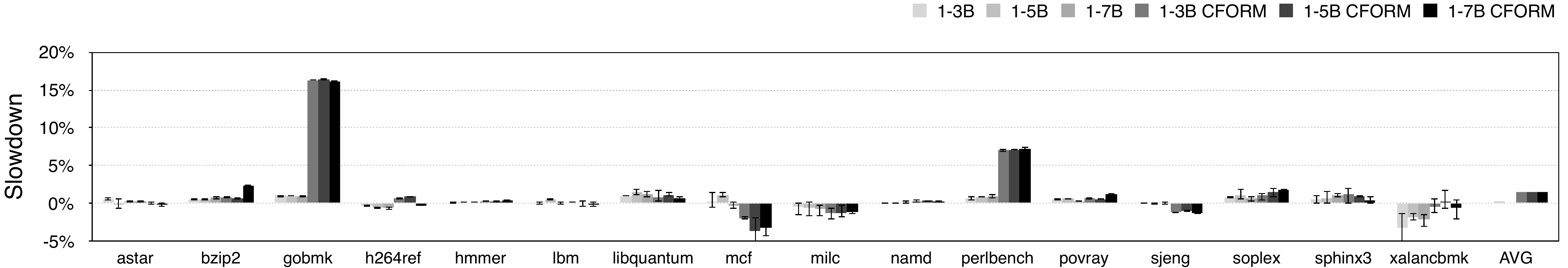}
  \caption{Slowdown of the intelligent insert policy with random sized \bytename{}s (with and without \INSTNAME{} instructions). The average slowdown is 2.0\%.}\label{fig:perf-random-intelligent-speccpu2006}
\end{figure*} %

Our evaluations so far revealed that the hardware modifications required to implement \Propname{} add little or no performance overhead.
Here, we evaluate the overheads incurred by the two software based changes required to enable intra-object memory safety with \Propname{}: the effect of underutilized memory structures (\eg{}caches) due to additional \bytename{}s, and the additional work necessary to issue \INSTNAME{} instructions (and the overhead of executing the instructions themselves).

\fakeitem\textbf{Evaluation Setup.}
We run the experiments on an Intel Skylake-based Xeon Gold 6126 processor running at 2.6GHz with RHEL Linux 7.5 (kernel 3.10).
We omit \texttt{dealII} and \texttt{omnetpp} since the shared libraries installed on RHEL are too old to execute these two \Propname{} enabled binaries, and \texttt{gcc} since it fails when executed with the memory allocator with inter-object spatial and temporal memory safety support.
The remaining 16 SPEC CPU2006 C/C++ benchmarks are compiled with our modified Clang version 6.0.0 with ``\texttt{-O3 -fno-strict-aliasing}'' flags.
We use the \texttt{ref} inputs and run to completion.
We run each benchmark-input pair five times and use the shortest execution time as its performance.
For the benchmarks with multiple \texttt{ref} inputs, the sum of the execution time of all the inputs are used as their execution times.\footnote{We use the arithmetic mean of the speedup (execution time of the original system divided by that of the system with additional latency) to compute the average, or in other words, we are interested in a condition where the workloads are not fixed and all types of workloads are equally probable on the target system\Cite{Eeckhout:2010wh,John:2004wg}.}

We estimate the performance impact of executing a \INSTNAME{} instruction by emulating it with a dummy store instruction that writes some value to the corresponding cache line's padding byte.
Since one \INSTNAME{} instruction can \prop{} the entire cache line, issuing one dummy store instruction per to-be-\proped{} cache line suffices.
In order to issue the dummy stores, we implement a LLVM pass to instrument the code to hook into memory allocations and deallocations.
We then retrieve the type information to locate the padding bytes, calculate the number of dummy stores and the address they access, and finally emit them.
Therefore, all the software overheads we need to pay to enable \Propname{} are accounted for in our evaluation.

For the random sized \bytename{}s, we evaluate three variants: we fix the minimum size to one byte while varying the maximum size to three, five and seven bytes (\ie{}on average the amount of \bytename{}s inserted are two, three and four bytes, respectively).
In addition, in order to account for the randomness introduced by the compiler, we generate three different versions of binaries for the same setup (\eg{}three versions of \texttt{astar} with random sized paddings of minimum one byte and maximum three bytes).
The error bars in the figure represent the minimum and the maximum execution times among 15 executions (three binaries~$\times$~five runs) and the average of the execution times is represented as the bar.

\fakeitem\textbf{Performance of the Opportunistic and Full Insertion Policies with \INSTNAME{} Instructions.}
\autoref{fig:perf-random-full-speccpu2006} presents the slowdown incurred by three set of strategies: full insertion policy (with random sized \bytename{}s) \emph{without} \INSTNAME{} instructions, the opportunistic policy \emph{with} \INSTNAME{} instructions, and the full insertion policy \emph{with} \INSTNAME{} instructions.
Since the first strategy does not execute \INSTNAME{} instructions it does not offer any security coverage, but is shown as a reference to showcase the performance breakdown of the third strategy (cache underutilization vs.\ executing \INSTNAME{} instructions).

First, we focus on the three variants of the first strategy, which is shown in the three left most bars.
We can see that different sizes of random sized \bytename{}s does not make a large difference in terms of performance.
The average slowdown of the three variants for the policy without \INSTNAME{} instructions are 5.5\%, 5.6\% and 6.5\%, respectively.
This can be backed up by our results shown in \autoref{fig:avg-perf-fixed-full-speccpu2006}, where the average slowdowns of additional padding of two, three and four bytes ranges from 5.4\% to 6.2\%.
Therefore in order to achieve higher security coverage without losing performance, using a random sized bytes of, minimum of one byte and maximum of seven bytes, is promising.
When we focus on individual benchmarks, we can see that a few benchmarks including \texttt{h264ref}, \texttt{mcf}, \texttt{milc} and \texttt{omnetpp} incur noticeable slowdowns (ranging from 15.4\% to 24.3\%).

Next, we examine the opportunistic policy \emph{with} \INSTNAME{} instructions, which is shown in the middle (fourth) bar.
Since this strategy does not add any additional \bytename{}s, the overheads are purely due to the work required to setup and execute \INSTNAME{} instructions.
The average slowdown of this policy is 7.9\%.
There are benchmarks which encounter a slowdown of more than 10\%, namely \texttt{gobmk}, \texttt{h264ref} and \texttt{perlbench}.
The overheads are due to frequent allocations and deallocations made during program execution, where the programs have to calculate and execute \INSTNAME{} instructions upon every event (since every compound data type will be/was \proped{}).
For instance \texttt{perlbench} is notorious for being malloc-intensive, and reported as such elsewhere\Cite{Serebryany:2012wl}.

Lastly the third policy, the full insertion policy \emph{with} \INSTNAME{} instructions, offers the highest security coverage in \Propname{} based system with the highest average slowdown of 14.0\% (with the random sized \bytename{}s of maximum seven bytes).
Nearly half (seven out of 16) the benchmarks encounter a slowdown of more than 10\%, which might not be suitable for performance-critical environments, and thus the user might want to consider the use of the following intelligent insertion policy.

\fakeitem\textbf{Performance of the Intelligent Insertion Policy with \INSTNAME{} Instructions.}
\autoref{fig:perf-random-intelligent-speccpu2006} shows the slowdowns of the intelligent insertion policy with random sized \bytename{}s (\emph{with} and \emph{without} \INSTNAME{} instructions, in the same spirit as \autoref{fig:perf-random-full-speccpu2006}).
First we focus on the strategy without executing \INSTNAME{} instructions (the three bars on the left).
The performance trend is similar such that the three variants with different random sizes have little performance difference, where the average slowdown is 0.2\% with the random sized \bytename{}s of maximum seven bytes.
We can see that none of the programs incurs a slowdown of greater than 5\%.
Finally with \INSTNAME{} instructions (three bars on the right), \texttt{gobmk} and \texttt{perlbench} have slowdowns of greater than 5\% (16.1\% for \texttt{gobmk} and 7.2\% for \texttt{perlbench}).
The average slowdown is 1.5\%, where considering its security coverage and performance overheads the intelligent policy might be the most practical option for many environments.

\section{Related Work}\label{sec:relatedwork}

\begin{figure}[!t] %
  \centering
  \subcaptionbox{Disjoint metadata whitelisting.\label{fig:prevworks-a}}{\includegraphics[height=0.45\columnwidth]{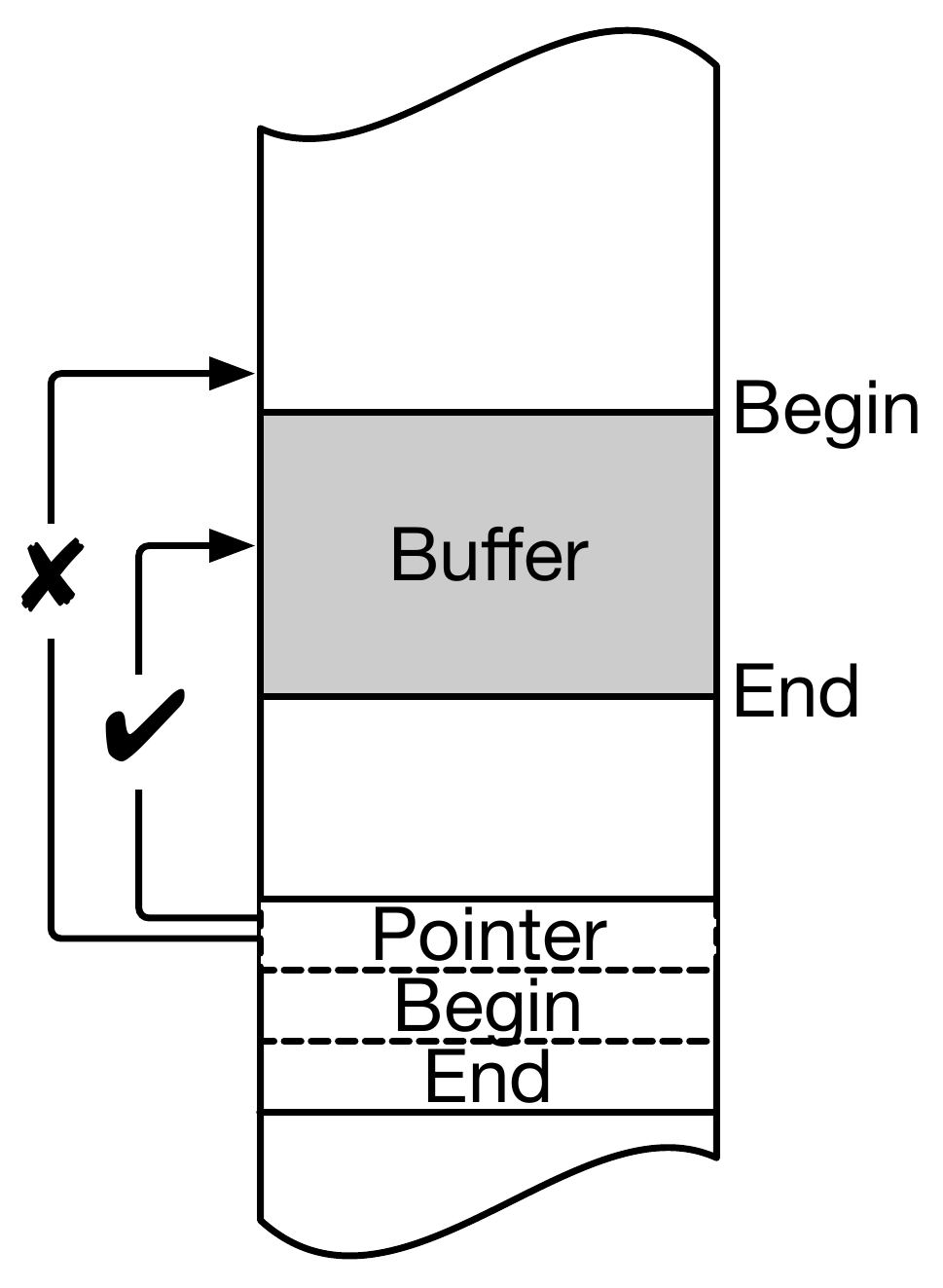}}
  \hspace{0mm}
  \subcaptionbox{Cojoined metadata whitelisting.\label{fig:prevworks-b}}{\includegraphics[height=0.45\columnwidth]{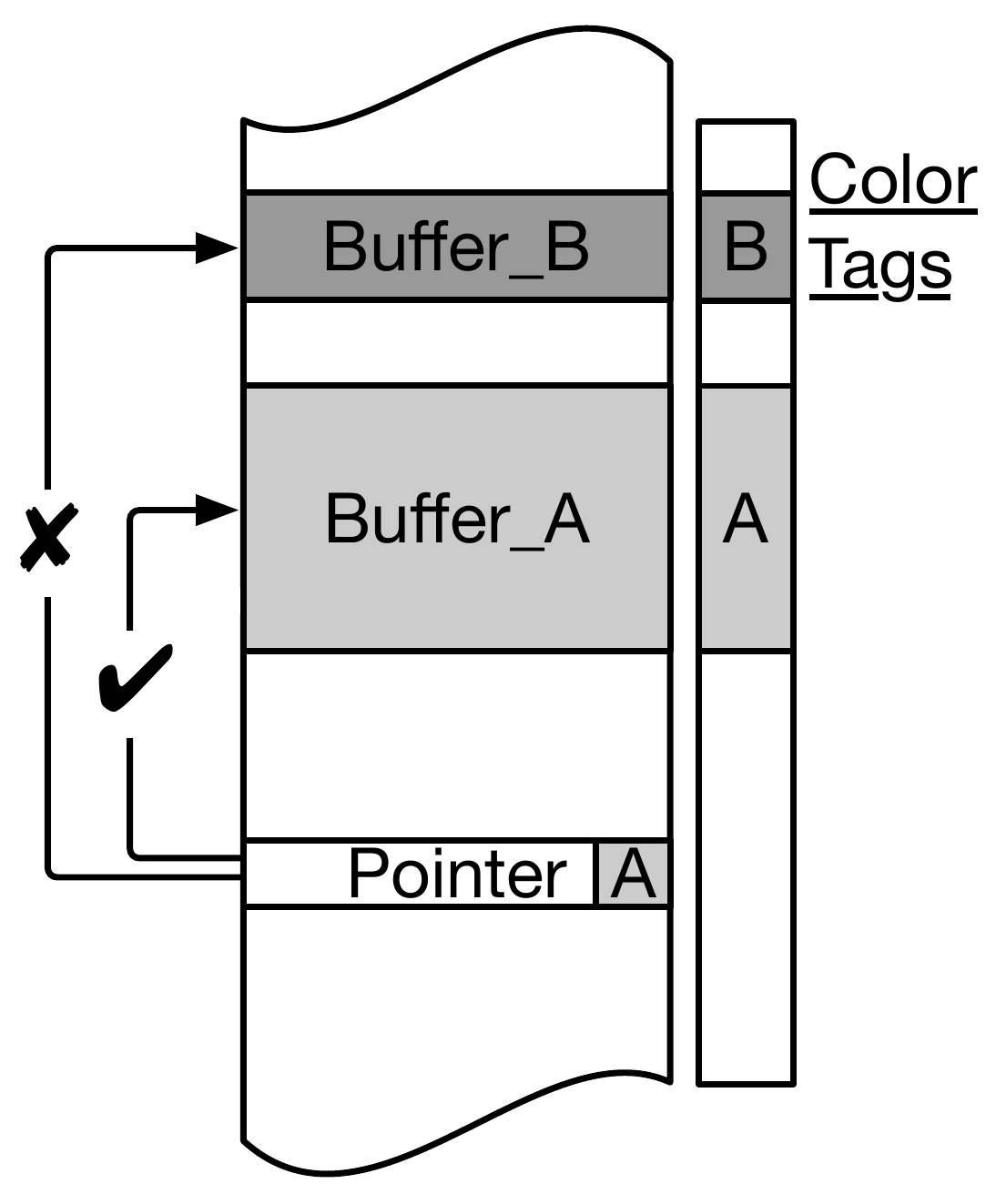}}
  \hspace{0mm}
  \subcaptionbox{Inlined metadata blacklisting.\label{fig:prevworks-c}}{\includegraphics[height=0.45\columnwidth]{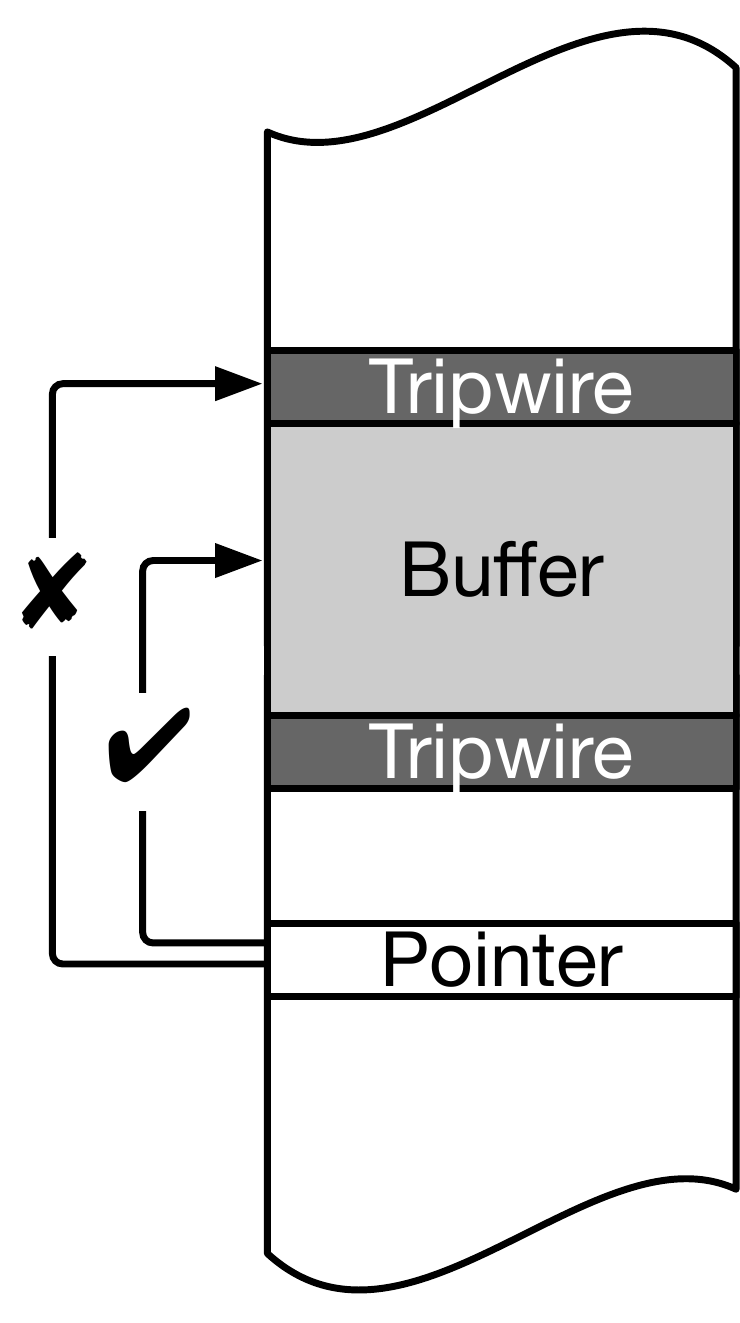}}
  \caption{Three main classes of hardware solutions for memory safety.}\label{fig:prevworks}
\end{figure}

Implementations of various safety mechanisms in hardware were very popular in the 70--90s, introducing crucial legacy techniques such as capabilities, segmentation and virtual memory.
Subsequently, the focus shifted towards scalability and performance until the last decade, when security saw a revival in interest.
In this section, we only focus on the latter group of modern hardware based security techniques, and compare them to \Propname{}.
Previous hardware solutions in this domain can be broadly categorized into the following three classes: disjoint metadata whitelisting, cojoined metadata whitelisting and inlined metadata blacklisting, as presented in \autoref{fig:prevworks}.

\fakeitem\textbf{Disjoint Metadata Whitelisting.} This class of techniques, also called base and bounds, attaches bounds metadata with every pointer, bounding the region of memory they can legitimately dereference (see \autoref{fig:prevworks-a}).
Hardbound\Cite{Devietti:2008kw} was the first hardware proposal to provide spatial memory safety using this mechanism.
Intel MPX\Cite{Oleksenko:2018kz} is similar, but also introduces explicit architectural interface (registers and instructions) for managing bounds information.
Temporal safety was introduced to this scheme by storing an additional ``version'' information along with the pointer metadata and verifying that no stale versions are ever retrieved\Cite{Nagarakatte:2012uf,Nagarakatte:2014cn}.
BOGO\Cite{Zhang:2019kp} adds temporal safety to MPX by invalidating all pointers to freed regions in MPX's lookup table.
Introduced about 35 years ago in commercial chips like Intel 432 and IBM System/38, the CHERI\Cite{Woodruff:2014tn} revived capability based architectures.
It has similar bounds-checking guarantees, in addition to having other metadata fields pertaining to permissions, etc\footnote{A recent version of CHERI\Cite{Woodruff:2019jk}, however, manages to compress metadata to 128 bits and change pointer layout to store it with the pointer value (\ie{}implementing base and bounds as cojoined metadata whitelisting), accordingly introducing instructions to manipulate them specifically.}.
PUMP\Cite{Dhawan:2015kv}, on the other hand, is a general-purpose framework for metadata propagation, and can be used for propagating pointer bounds.

Typically, per pointer metadata is stored separately from the pointer in a shadow memory region, in order to maintain legacy pointer layout assumptions.
Thus, although metadata storage overhead scales according to the number of pointers in principle, techniques generally reserve a fixed chunk of memory for easy lookup.
Owing to this disjoint nature, metadata access therefore requires additional memory operations, which individual proposals seek to minimize with caching and other optimizations.
Regardless, disjoint metadata introduces atomicity concerns potentially resulting in false positives and negatives or complicating coherence designs at the least (\eg{}MPX is not thread-safe).
Explicit specification of bounds per pointer also allows bounds-narrowing in principle, wherein pointer bounds can be tailored to protect individual elements in a composite memory object (for instance, when passing the pointer to an element to another function).
However, commercial compilers do not support this feature for MPX due to the complexity of compiler analyses required.
Furthermore, compatibility issues with untreated modules (unprotected libraries, for instance) also introduces real-world deployability concerns for these techniques.
For instance MPX drops its bounds when protected pointers are modified by unprotected modules, while CHERI does not support it at all.
MPX additionally makes bounds checking explicit, thus introducing a marginal computational overhead to bounds management as well.

\fakeitem\textbf{Cojoined Metadata Whitelisting.} Originally introduced in the IBM System/360 mainframes, this mechanism assigns a ``color'' to memory chunks as well as pointers.
As such, the runtime check for access validity simply consists of comparing the colors of the pointer and accessed memory (see \autoref{fig:prevworks-b}).

This technique is currently commercially deployed by SPARC ADI\Cite{ADI:2015ha},\footnote{ARM has a similar upcoming Memory Tagging\Cite{Arm:Tag} feature, whose implementation details are unclear, as of this work.} which refactors unused higher order bits in pointers to store the color.
Color associated with memory is stored in the ECC bits while in memory, and dedicated per line metadata bits while in cache.
Due to the latter feature, metadata storage does not occupy any additional memory in the program's address space.\footnote{When a memory is swapped, color bits are copied into memory by the OS however.}
Additionally, since metadata bits are acquired along with concomitant data, extra memory operations are obviated.
For the same reason, it is also compatible with unprotected modules since the checks are implicit as well.
Temporal safety is trivially achieved by assigning a different color when memory regions are reused.
However, intra-object protection or bounds-narrowing is not supported as there is no means for ``overlapping'' colors.
Furthermore, protection is also dependent on the number of metadata bits employed, since it determines the number of colors that can be assigned.
So, while color reuse allows ADI to scale and limit metadata storage overhead, it can also be exploited by this vector.
Another disadvantage of this technique, specifically due to inlining metadata in pointers, is that it only supports 64-bit architectures.
Narrower pointers would not have enough spare bits to accommodate color information.

\begin{table}[t]
  \centering
  \footnotesize
  \resizebox{1.00\columnwidth}{!}{
  \begin{tabular}{r r l l l}
    \toprule
    \multicolumn{1}{c}{\multirow{2}{*}{\textbf{Proposal}} } & \multicolumn{1}{c}{\textbf{Protection}}  & \multicolumn{1}{c}{\textbf{Intra-}}  & \multicolumn{1}{c}{\textbf{Binary}}        & \multicolumn{1}{c}{\textbf{Temporal}} \\
                                                            & \multicolumn{1}{c}{\textbf{Granularity}} & \multicolumn{1}{c}{\textbf{Object}}  & \multicolumn{1}{c}{\textbf{Composability}} & \multicolumn{1}{c}{\textbf{Safety}} \\
    \midrule
    Hardbound\Cite{Devietti:2008kw}                         & Byte                                     & \hspace{3mm}\cmark{}$^\ast$          & \hspace{8mm}\xmark{}                       & \hspace{5mm}\xmark{} \\
    Watchdog\Cite{Nagarakatte:2012uf}                       & Byte                                     & \hspace{3mm}\cmark{}$^\ast$          & \hspace{8mm}\xmark{}                       & \hspace{5mm}\cmark{} \\
    WatchdogLite\Cite{Nagarakatte:2014cn}                   & Byte                                     & \hspace{3mm}\cmark{}$^\ast$          & \hspace{8mm}\xmark{}                       & \hspace{5mm}\cmark{} \\
    Intel MPX\Cite{Oleksenko:2018kz}                        & Byte                                     & \hspace{3mm}\cmark{}$^\ast$          & \hspace{8mm}\xmark{}$^\ddagger$            & \hspace{5mm}\xmark{} \\
    BOGO\Cite{Zhang:2019kp}                                 & Byte                                     & \hspace{3mm}\cmark{}$^\ast$          & \hspace{8mm}\xmark{}$^\ddagger$            & \hspace{5mm}\cmark{} \\
    PUMP\Cite{Dhawan:2015kv}                                & Word                                     & \hspace{3mm}\xmark{}                 & \hspace{8mm}\cmark{}                       & \hspace{5mm}\cmark{} \\
    CHERI\Cite{Woodruff:2014tn}                             & Byte                                     & \hspace{3mm}\xmark{}$^\dagger$       & \hspace{8mm}\xmark{}                       & \hspace{5mm}\xmark{} \\
    CHERI concentrate\Cite{Woodruff:2019jk}                 & Byte                                     & \hspace{3mm}\xmark{}$^\dagger$       & \hspace{8mm}\xmark{}                       & \hspace{5mm}\xmark{} \\
    SPARC ADI\Cite{ADI:2015ha}                              & Cache line                               & \hspace{3mm}\xmark{}                 & \hspace{8mm}\cmark{}                       & \hspace{5mm}\cmark{}$^\mathsection$ \\
    SafeMem\Cite{Qin:2005dm}                                & Cache line                               & \hspace{3mm}\xmark{}                 & \hspace{8mm}\cmark{}                       & \hspace{5mm}\xmark{} \\
    REST\Cite{Sinha:2018eh}                                 & 8--64B                                   & \hspace{3mm}\xmark{}                 & \hspace{8mm}\cmark{}                       & \hspace{5mm}\cmark{}$^\mathparagraph$ \\
    \cmidrule{1-5}
    \textbf{\Propname{}}                                    & Byte                                     & \hspace{3mm}\cmark{}                 & \hspace{8mm}\cmark{}                       & \hspace{5mm}\cmark{}$^\mathparagraph$\\
    \cmidrule{1-5}
  \end{tabular}
  }
  \caption{Security comparison against previous hardware techniques. $^\ast$Achieved with bounds narrowing. $^\dagger$Although the hardware supports bounds narrowing, CHERI foregoes it since doing so compromises capability
  logic\Cite{Davis:2019hb}$.^\ddagger$Execution compatible, but protection dropped when external modules modify pointer. $^\mathsection$Limited to 13 tags. $^\mathparagraph$Allocator should randomize allocation predictability.}\label{tab:sec_comparison}
\end{table}

{
\makeatletter
\def\adl@drawiv#1#2#3{%
        \hskip.5\tabcolsep
        \xleaders#3{#2.5\@tempdimb #1{1}#2.5\@tempdimb}%
                #2\z@ plus1fil minus1fil\relax
        \hskip.5\tabcolsep}
\newcommand{\cdashlinelr}[1]{%
  \noalign{\vskip\aboverulesep
           \global\let\@dashdrawstore\adl@draw
           \global\let\adl@draw\adl@drawiv}
  \cdashline{#1}
  \noalign{\global\let\adl@draw\@dashdrawstore
           \vskip\belowrulesep}}
\makeatother
\newcommand{\appropto}{\mathrel{\vcenter{
  \offinterlineskip\halign{\hfil$##$\cr
    \propto\cr\noalign{\kern2pt}\sim\cr\noalign{\kern-2pt}}}}}
\renewcommand{\arraystretch}{1.00}

\begin{table*}[t]
  \centering
  \resizebox{1.00\textwidth}{!}{
  \begin{tabular}{r r l l r}
    \toprule
    \multicolumn{1}{c}{\multirow{2}{*}{\textbf{Proposal}} } & \multicolumn{1}{c}{\textbf{Metadata}} & \multicolumn{1}{c}{\textbf{Memory}}                               & \multicolumn{1}{c}{\textbf{Performance}} & \multicolumn{1}{c}{\textbf{Main}} \\
                                                            & \multicolumn{1}{c}{\textbf{Overhead}} & \multicolumn{1}{c}{\textbf{Overhead}}                             & \multicolumn{1}{c}{\textbf{Overhead}}    & \multicolumn{1}{c}{\textbf{Operations}} \\
    \midrule
    \multirow{2}{*}{Hardbound\Cite{Devietti:2008kw}}        & 0--2 words per ptr,                   & \multirow{2}{*}{$\appropto$ \# of ptrs and prog memory footprint} & \multirow{2}{*}{$\appropto$ \# of ptr derefs}            & 1--2 mem ref for bounds (may be cached), \\
                                                            & 4b per word                           &                                                                   &                                                          & check $\mu$ops. \\
    \cdashlinelr{1-5}
    \multirow{2}{*}{Watchdog\Cite{Nagarakatte:2012uf}}      & \multirow{2}{*}{4 words per ptr}      & \multirow{2}{*}{$\appropto$ \# of ptrs and allocations}           & \multirow{2}{*}{$\appropto$ \# of ptr derefs}            & 1--3 mem ref for bounds (may be cached), \\
                                                            &                                       &                                                                   &                                                          & check $\mu$ops. \\
    \cdashlinelr{1-5}
    \multirow{2}{*}{WatchdogLite\Cite{Nagarakatte:2014cn}}  & \multirow{2}{*}{4 words per ptr}      & \multirow{2}{*}{$\appropto$ \# of ptrs and allocations}           & \multirow{2}{*}{$\appropto$ \# of ptr ops}               & 1--3 mem ref for bounds (may be cached), \\
                                                            &                                       &                                                                   &                                                          & check \& propagate insns. \\
    \cdashlinelr{1-5}
    \multirow{2}{*}{Intel MPX\Cite{Oleksenko:2018kz}}       & \multirow{2}{*}{2 words per ptr}      & \multirow{2}{*}{$\appropto$ \# of ptrs}                           & \multirow{2}{*}{$\appropto$ \# of ptr derefs}            & 2+ mem ref for bounds (may be cached), \\
                                                            &                                       &                                                                   &                                                          & check \& propagate insns. \\
    \cdashlinelr{1-5}
    \multirow{2}{*}{BOGO\Cite{Zhang:2019kp}}                & \multirow{2}{*}{2 words per ptr}      & \multirow{2}{*}{$\appropto$ \# of ptrs}                           & \multirow{2}{*}{$\appropto$ \# of ptr derefs}            & MPX ops + ptr miss exception handling, \\
                                                            &                                       &                                                                   &                                                          & page permission mods. \\
    \cdashlinelr{1-5}
    \multirow{2}{*}{PUMP\Cite{Dhawan:2015kv}}               & \multirow{2}{*}{64b per cache line}   & \multirow{2}{*}{$\appropto$ Prog memory footprint}                & \multirow{2}{*}{$\appropto$ \# of ptr ops}               & 1 mem ref for tags, may be cached, \\
                                                            &                                       &                                                                   &                                                          & fetch and chk rules; propagate tags. \\
    \cdashlinelr{1-5}
    \multirow{2}{*}{CHERI\Cite{Woodruff:2014tn}}            & \multirow{2}{*}{256b per ptr}         & \multirow{2}{*}{$\appropto$ \# of ptrs and physical mem}          & \multirow{2}{*}{$\appropto$ \# of ptr ops}               & 1+ mem ref for capability (may be cached), \\
                                                            &                                       &                                                                   &                                                          & capability management insns. \\
    \cdashlinelr{1-5}
    \multirow{2}{*}{CHERI concentrate\Cite{Woodruff:2019jk}}& \multirow{2}{*}{Ptr size is 2x}       & \multirow{2}{*}{$\appropto$ \# of ptrs}                           & \multirow{2}{*}{$\appropto$ \# of ptr ops}               & Wide ptr load (may be cached), \\
                                                            &                                       &                                                                   &                                          & capability management insns. \\
    \cdashlinelr{1-5}
    SPARC ADI\Cite{ADI:2015ha}                              & 4b per cache line                     & $\appropto$ Prog memory footprint                                 & $\appropto$ \# of tag (un)set ops        & (Un)set tag. \\
    \cdashlinelr{1-5}
    SafeMem\Cite{Qin:2005dm}                                & 2x blacklisted memory                 & $\appropto$ Blacklisted memory                                    & $\appropto$ \# of ECC (un)set ops        & Syscall to scramble ECC, copy data content. \\
    \cdashlinelr{1-5}
    REST\Cite{Sinha:2018eh}                                 & 8--64B token                          & $\appropto$ Blacklisted memory                                    & $\appropto$ \# of arm/disarm insns       & Execute arm/disarm insns. \\
    \cmidrule{1-5}
    \textbf{\Propname{}}                                    & Byte granular \bytename{}             & $\appropto$ Blacklisted memory                                    & $\appropto$ \# of \INSTNAME{} insns.     & Execute \INSTNAME{} insns. \\
    \cmidrule{1-5}
  \end{tabular}
    }
  \caption{Performance comparison against previous hardware techniques.}\label{tab:perf_comparison}
\end{table*}

\begin{table*}[t]
  \centering
  \resizebox{1.00\textwidth}{!}{
  \begin{tabular}{r r r r r}
    \toprule
    \multicolumn{1}{c}{\textbf{Proposal}}                   & \multicolumn{1}{c}{\textbf{Core}}             & \multicolumn{1}{c}{\textbf{Caches/TLB}} & \multicolumn{1}{c}{\textbf{Memory}} & \multicolumn{1}{c}{\textbf{Software}} \\
    \midrule
    \multirow{3}{*}{Hardbound\Cite{Devietti:2008kw}}        & $\mu$op injection \& logic for ptr meta,      & \multirow{3}{*}{Tag cache and its TLB}     & \multirow{3}{*}{N/A} & \multirow{3}{*}{Compiler \& allocator annotates ptr meta} \\
                                                            & extend reg file and data path to              &                          &     & \\
                                                            & propagate ptr meta                            &                          &     & \\
    \cdashlinelr{1-5}
    \multirow{3}{*}{Watchdog\Cite{Nagarakatte:2012uf}}      & $\mu$op injection \& logic for ptr meta,      & \multirow{3}{*}{Ptr lock cache}           & \multirow{3}{*}{N/A} & \multirow{3}{*}{Compiler \& allocator annotates ptr meta}  \\
                                                            & extend reg file and data path to              &                          &     & \\
                                                            & propagate ptr meta                            &                          &     & \\
    \cdashlinelr{1-5}
    \multirow{2}{*}{WatchdogLite\Cite{Nagarakatte:2014cn}}  & \multirow{2}{*}{N/A}                          & \multirow{2}{*}{N/A}     & \multirow{2}{*}{N/A} & Compiler \& allocator annotates ptrs, \\
                                                            &                                               &                          &                      & compiler inserts meta propagation and check insns \\
    \cdashlinelr{1-5}
    \multirow{2}{*}{Intel MPX\Cite{Oleksenko:2018kz}}       & \multicolumn{3}{c}{\multirow{2}{*}{Unknown (closed platform\Cite{MPX:2018me}, design likely similar to Hardbound)}} & Compiler \& allocator annotates ptrs, \\
                                                            &                                               &                          &        & compiler inserts meta propagation and check insns \\
    \cdashlinelr{1-5}
    \multirow{2}{*}{BOGO\Cite{Zhang:2019kp}}                & \multicolumn{3}{c}{\multirow{2}{*}{Unknown (closed platform\Cite{MPX:2018me}, design likely similar to Hardbound)}} & MPX mods + kernel mods for bounds page \\
                                                            &                                               &                          &     & right management \\
    \cdashlinelr{1-5}
    \multirow{3}{*}{PUMP\Cite{Dhawan:2015kv}}               & Extend all data units by tag width,           & \multirow{3}{*}{Rule cache} & \multirow{3}{*}{N/A} & \multirow{3}{*}{Compiler \& allocator (un)sets memory, tag ptrs} \\
                                                            & modify pipeline stages for tag checks,        &                                &                       & \\
                                                            & new miss handler                              &                                &                       & \\
    \cdashlinelr{1-5}
    \multirow{2}{*}{CHERI\Cite{Woodruff:2014tn}}            & Capability reg file, coprocessor              & \multirow{2}{*}{Capability caches}        & \multirow{2}{*}{N/A} & Compiler \& allocator annotates ptrs, \\
                                                            & integrated with pipeline                      &                          &     & compiler inserts meta propagation and check insns \\
    \cdashlinelr{1-5}
    \multirow{2}{*}{CHERI concentrate\Cite{Woodruff:2019jk}}& \multirow{2}{*}{Modify pipeline to integrate ptr checks}       & \multirow{2}{*}{N/A}  & \multirow{2}{*}{N/A} & Compiler \& allocator annotates ptrs, \\
                                                            &                                               &                          &     & compiler inserts meta propagation and check insns \\
    \cdashlinelr{1-5}
    SPARC ADI\Cite{ADI:2015ha}                              & \multicolumn{3}{c}{Unknown (closed platform)}                                                         & Compiler \& allocator (un)sets memory, tag ptrs \\
    \cdashlinelr{1-5}
    SafeMem\Cite{Qin:2005dm}                                & N/A                                           & N/A                      & Repurposes ECC bits &  \\
    \cdashlinelr{1-5}
    \multirow{2}{*}{REST\Cite{Sinha:2018eh}}                & \multirow{2}{*}{N/A}                          & 1--8b per L1D line,         & \multirow{2}{*}{N/A} & Compiler \& allocator (un)sets tags, \\
                                                            &                                               & 1 comparator             &     & allocator randomizes allocation order/placement \\
    \cmidrule{1-5}
    \multirow{2}{*}{\textbf{\Propname{}}}                   & \multirow{2}{*}{N/A}                          & 8b per L1D line,  & \multirow{2}{*}{Use unused ECC bits} & Compiler \& allocator mods to (un)set tags, \\
                                                            &                                               & 1b per L2/L3 line &                     & compiler inserts intra-object spacing \\
    \cmidrule{1-5}
  \end{tabular}
  }
  \caption{Comparison of implementation complexity among previous hardware techniques.}\label{tab:impl_comparison}
\end{table*}
}

\fakeitem\textbf{Inlined Metadata Blacklisting.} Another line of work, also referred to as tripwires, aims to detect overflows by simply blacklisting a patch of memory on either side of a buffer, and flagging accesses to this patch (see \autoref{fig:prevworks-c}).
This is very similar to contemporary canary design\Cite{Cowan:1998wc}, but there are a few critical differences.
First, canaries only detect overwrites, not overreads.
Second, hardware tripwires trigger instantaneously, whereas canaries need to be periodically checked for integrity, providing a period of attack to time of use window.
Finally, unlike hardware tripwires, canary values can be leaked or tampered, and thus mimicked.

SafeMem\Cite{Qin:2005dm} implements tripwires by repurposing ECC bits in memory to mark memory regions invalid, thus trading off reliability for security.
On processors supporting speculative execution, however, it might be possible to speculatively fetch blacklisted lines into the cache without triggering a faulty memory exception.
Unless these lines are flushed immediately after, SafeMem's blacklisting feature can be trivially bypassed.
Alternatively, REST\Cite{Sinha:2018eh} achieves the same by storing a predetermined large random number, in the form of a 64B tokens, in the memory to be blacklisted.
Violations are detected by comparing cache lines with the token when they are fetched.
REST provides temporal safety by quarantining freed memory, and not reusing them for subsequent allocations.
Compatibility with unprotected modules is easily achieved as well, since tokens are part of the program's address space and all access are implicitly checked.
However, intra-object safety was not supported by REST owing to the large memory overhead such heavy usage of tokens would entail.

Since it operates on the principle of detecting memory accesses to \bytename{}s, which are in turn stored along with program data, \Propname{} belongs to the inlined metadata class of defenses.
However, it is different from other works in the class in one key aspect --- granularity.
While both REST and SafeMem blacklisted at the cache line granularity, \Propname{} does so at the byte granularity.
It is this property that enables us to provide intra-object safety with negligible performance and memory overheads, unlike previous work in the area.
For inter-object spatial safety and temporal safety, we employ the same design principles as REST.\@
Hence, our safety guarantees are a \textit{strict superset} of those provided by previous schemes in this class (spatial safety by blacklisting and temporal safety by quarantining).

\subsection{Comparison with \Propname{}}

Tables~\ref{tab:sec_comparison},~\ref{tab:perf_comparison}, and~\ref{tab:impl_comparison} summarize the performance, security, and implementation characteristics of the hardware based memory safety techniques discussed in this section respectively.
\Propname{} has the advantage of requiring simpler hardware modifications and being faster than disjoint metadata based whitelisting systems.
The hardware savings mainly stem from the fact that our metadata resides with program data; it does not require explicit propagation while additionally obviating all lookup logic.
This significantly reduces our design's implementation costs.
\Propname{} also has lower performance and energy overheads since it neither requires multiple memory accesses, nor does it incur any significant checking costs.
However, unlike them, \Propname{} can be bypassed if accesses to \bytename{}s can be avoided (further discussed in \autoref{sec:discussion}).
This safety-vs.-complexity tradeoff is critical to deployability and we argue that our design point is more practical. This is because designers have to contend with integrating these features to already complicated processor designs, without introducing additional bugs while also keeping the functionality of legacy software intact.
This is a hard balance to strike\Cite{Oleksenko:2018kz}.

On the other hand, ideal cojoined metadata mechanisms would have comparable slowdowns and similar compiler requirements.
However practical implementations like ADI exhibits some crucial differences from the ideal.
\begin{itemize}
    \item It is limited to 64-bit architectures, which excludes a large portion of embedded and IoT processors that operate on 32-bit or narrower platforms.
    \item It has finite number of colors since available tag bits are limited --- ADI supports 13 colors with 4 tag bits. This is important because reusing colors proportionally reduces the safety guarantees of these systems in the event of a collision.
    \item It operates at the coarse granularity of cache line width, and hence, is not practically applicable for intra-object safety.
\end{itemize}

On the contrary, \Propname{} is agnostic of architecture width and is, hence, better suited for deployment over a more diverse device environment.
In terms of safety, collision is not an issue for our design either.
Hence, unlike cojoined metadata systems, our security does not scale inversely with the number of allocations in the program (see \autoref{sec:discussion} for a detailed discussion).
Finally, our fine-grained protection also makes us suitable for intra-object memory safety which is a non-trivial threat in modern security\Cite{Lu:2016cu}.

\section{Conclusion}\label{sec:conclusion}

\Propname{} is a hardware primitive which allows blacklisting a memory location at byte granularity with low area and performance overhead.
A key observation behind \Propname{} is that a blacklisted region need not store useful data separately in most cases, since we can utilize byte-granular, existing or added, space present between object elements to store the metadata.
This in-place compact data structure also avoids additional operations for extraneously fetching the metadata making it very performant in comparison. Further, by changing how data is stored within a cache line we are able to reduce the hardware area overheads substantially.
Subsequently, if the processor accesses a \proped{} byte (or a \bytename{}), due to programming errors or malicious attempts, it reports a privileged exception.

To provide memory safety, we use \Propname{} to insert \bytename{}s within data structures (\eg{}between fields of a struct) upon memory allocation and clear them on deallocation.
Notably, by doing so, \Propname{} can even detect intra-object overflows, which is one of the prominent open problems in memory safety, despite decades of research in this area. We also described the necessary compiler and software support for providing memory safety using \Propname{}.
To the best of our knowledge, this is the first hardware primitive which makes in place byte-granular blacklisting practical.

{\bibliographystyle{unsrtnat}
\small
\bibliography{main}}

\renewcommand{\appendixpagename}{\Large{Appendices}}
\begin{appendices}
\section{\PROPNAME{} Variants}

Here we present two other variants of \propname{}-bitvector (designed for the L1 cache) which have less storage overhead (but with additional complexity) compared to the one presented in \autoref{subsec:califorms_bitvector}.

\begin{figure}[!t] %
  \centering
  \includegraphics[width=0.48\textwidth]{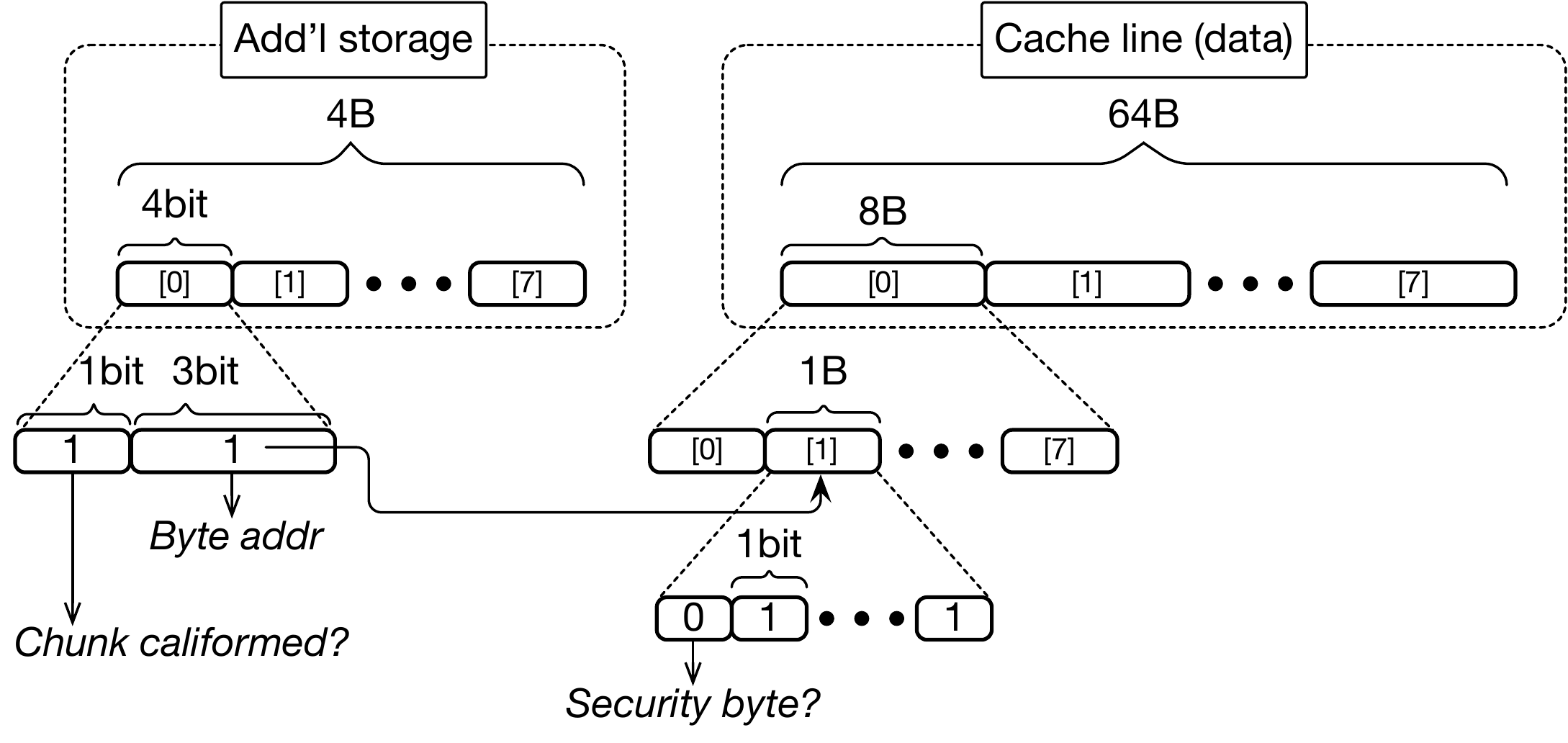}
  \caption{\Propname{}-bitvector that stores a bit vector inside \bytename{} locations. The additional metadata (4-bit per 8B) specifies if the corresponding chunk contains a \bytename{}, and if it does, where in the chunk the bit vector is stored in. HW overhead of 4B per 64B cache line.}\label{fig:califorms-bitvector-4byte}
\end{figure}

\fakeitem\textbf{\Propname{}-4B.}
The first variant has 4B of additional storage per 64B cache line.
This \Propname{} stores the bit vector \textit{within} a \bytename{} (illustrated in \autoref{fig:califorms-bitvector-4byte}).
Since a single byte bit vector (which can be stored in one \bytename{}) can represent the state for 8B of data, we divide the 64B cache line into eight 8B chunks.
If there is at least one \bytename{} within an 8B chunk, we use one of those bytes to store the bit vector which represents the state of the chunk.
For each chunk, we need to add four additional bits of storage.
One bit to represent whether the chunk is \textit{\proped{}} (contains a \bytename{}), and three bits to specify which byte within the chunk stores the bit vector.
Therefore, the additional storage is 4B (4-bit $\times$ 8 chunks) or 6.25\% per 64B cache line.
The figure highlights the chunk \texttt{[0]} being \proped{} where the corresponding bit vector is stored in byte \texttt{[1]}.

\begin{figure}[!t] %
  \centering
  \includegraphics[width=0.48\textwidth]{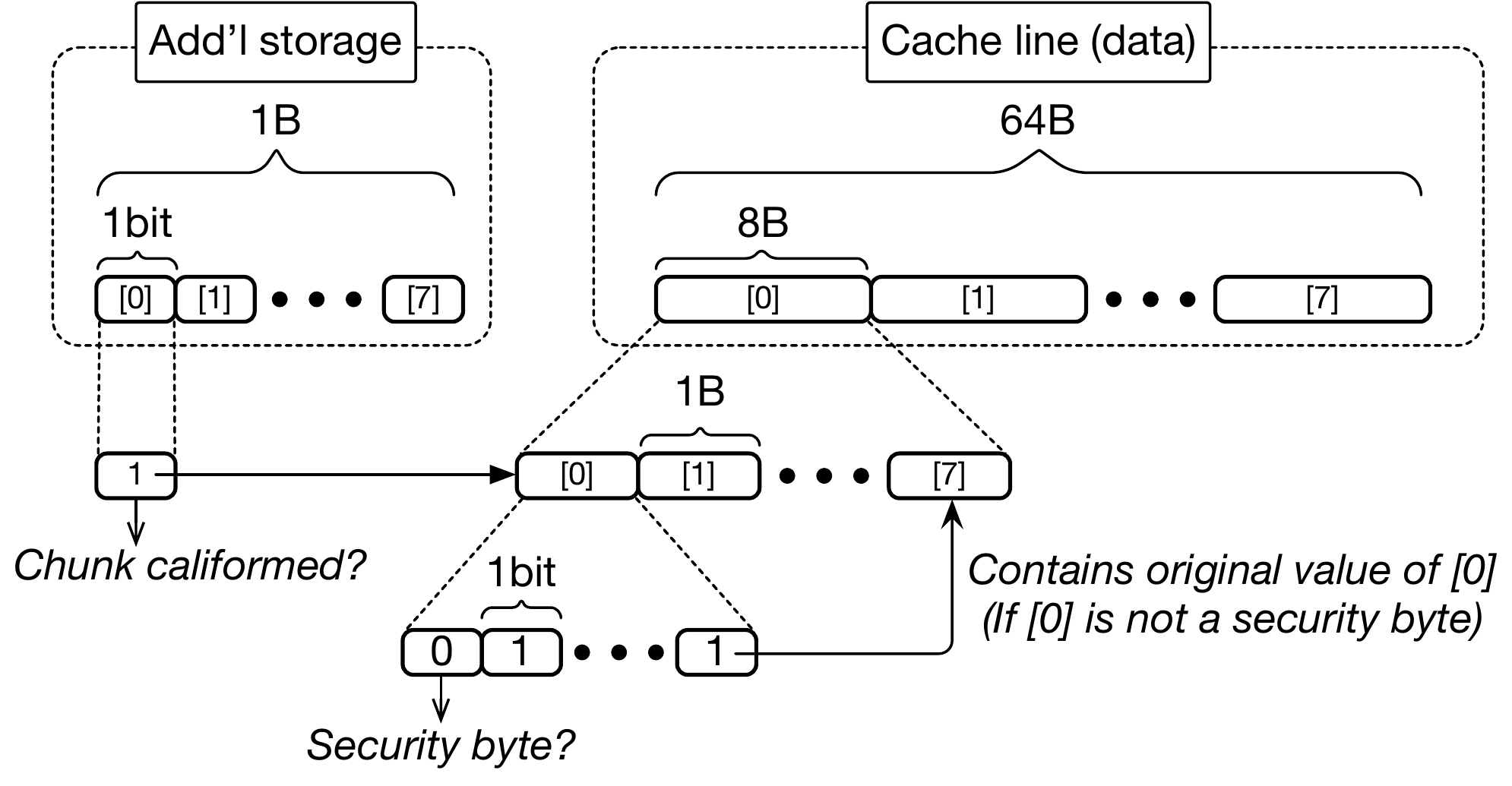}
  \caption{\Propname{}-bitvector that stores a bit vector in the header (0th) byte of the chunk. If the header byte is normal data (not a \bytename{}), its original value is stored in the last \bytename{}. The additional metadata (1-bit per 8B) specifies if the corresponding chunk contains a \bytename{}. HW overhead of 1B per 64B cache line.}\label{fig:califorms-bitvector-1byte}
\end{figure}

\fakeitem\textbf{\Propname{}-1B.}
We can further reduce the metadata overhead by restricting where we store the bit vector within the chunk (illustrated in \autoref{fig:califorms-bitvector-1byte}).
The idea is to always store the bit vector in a fixed location (the 0th byte in the figure, or the header byte; similar idea used in \propname{}-sentinel).
If the 0th byte is a \bytename{} this works without additional modification.
However if the 0th byte is \textit{not} a \bytename{}, we need to save its original value somewhere else so that we can retrieve it when required.
For this purpose, we use one of the \bytename{}s (the last \bytename{} is chosen in the figure).
This way we can eliminate three bits of metadata per chunk to address the byte which contains the bit vector.
Therefore, the additional storage is 1B or 1.56\% per 64B cache line.
Similar with \autoref{fig:califorms-bitvector-4byte}, the figure highlights the chunk \texttt{[0]} being \proped{} (where the corresponding bit vector is stored in the first byte) and the original value of byte \texttt{[0]} stored in the last \bytename{}, byte \texttt{[7]}, within the chunk.

\balance{}

\fakeitem\textbf{Evaluation.}
We perform the same VLSI evaluation shown in \autoref{subsec:evaluation_hardware} for the two additional \propname{}-bitvector introduced in this section.
\autoref{tab:vlsi_all} presents the results.
As we can see, \propname{}-bitvector with 4B and 1B overheads incur 47\% and 20\% of extra delay, respectively, upon L1 hit compared to the \propname{}-bitvector with 8B overhead (49\% and 22\% additional delay compared to the baseline L1 data cache without \Propname{}).
Also, both \propname{}-bitvector add almost the same overheads upon spill and fill operations (compared to the \propname{}-bitvector with 8B overhead) which are 9\% delay and 30\% energy for spill, and 34\% delay and 17\% energy for fill operations.
Our evaluation reveals that \propname{}-bitvector with 1B overhead outperforms the other with 4B overhead both in terms of additional storage and access latency/energy.
The reason is due to the design restriction of fixing the location of the header byte which allows faster lookup of the bit vector in the \bytename{}.
\Propname{}-bitvector with 1B overhead can be a good alternative (to the one presented in \autoref{sec:microarchitecture}) in domains where area budget is more tight and/or less performance critical; \eg{}embedded or IoT systems.

\begin{table*}[t!] %
  \caption{Area, delay and power overheads of the three L1 \Propname{} (GE represents gate equivalent). The top two rows are presented in \autoref{tab:vlsi} and are shown here again for reference. \Propname{}-bitvector with 4B and 1B overheads incur 47\% and 20\% extra delay, respectively, upon L1 hit compared to \propname{}-bitvector with 8B overhead. Also the two \Propname{} add 9\% delay and 30\% energy upon spill and 34\% delay and 17\% energy upon fill.\label{tab:vlsi_all}}{%
  \centering
  \resizebox{0.99\textwidth}{!}{
    \begin{tabular}{c  c c c  c c c  c c c  c c c}
      \toprule
Design & \multicolumn{3}{c}{Main synthesis results} & \multicolumn{3}{c}{L1 overheads} & \multicolumn{3}{c}{Fill overheads}  & \multicolumn{3}{c}{Spill overheads} \\
Name  & Area  (GE) & Delay  ($ns$)  & Power  ($m W$)    & Area  (\%)  & Delay(\%) & Power  (\%)   & Area  (GE) & Delay  ($ns$)  & Power  ($m W$) & Area  (GE) & Delay  ($ns$)  & Power  ($m W$) \\ %
      \midrule
Baseline & 347,329.19 & 1.62 & 15.84 & --- & --- & --- & --- & --- & --- & --- & --- & --- \\
      \midrule
\Propname{}-8B & 412,263.87 & 1.65 & 16.17 & 18.69 & 1.85 & 2.12 & 8,957.16 & 1.43 & 0.18 & 34,561.80 & 5.50 & 0.52  \\
      \midrule
\Propname{}-4B & 370,972.35 & 2.42 & 17.95 & 6.80 & 49.38 & 11.00 & 9,770.04 & 1.92 & 0.21 & 35,775.36 & 5.99 & 0.68      \\
      \midrule
\Propname{}-1B & 356,694.82 & 1.98 & 16.00 & 2.69 & 22.22 & 1.06 & 10,223.28 & 1.94 & 0.22 & 35,958.24 & 5.99 & 0.67 \\
      \bottomrule
    \end{tabular}
  }
}
\end{table*} %

\section{Handling SIMD/Vector Instructions}

As we have discussed in \autoref{subsec:object_type_layouts}, precise loads and stores (along with the whitelisting capability) allow us to detect access violation upon memory instructions.
However, there are certain class of instructions where issuing precise memory instructions may noticeably degrade performance.
One class we can imagine is the SIMD/vector instructions where vector loads read a very wide (\eg{}512 bits for Intel AVX-512) word into the SIMD/vector register with a single instruction.
For such instructions we can (1) operate the same way with regular loads by issuing precise loads (\eg{}by using vector gather instructions with appropriate masks), (2) issue wide vector loads as is and trigger an exception whenever the vector load touches a \bytename{}; this may introduce false positives but in reality data structures used by SIMD/vector instructions are unlikely to contain \bytename{}s, or (3) add one bit per byte in the SIMD/vector registers so that we can propagate the \bytename{} information, and trigger an exception whenever SIMD/vector instructions operates on a \bytename{}.
Investigating these alternatives are left for future work.

\end{appendices}

\end{document}